\documentclass[fleqn,usenatbib]{mnras}

\usepackage{newtxtext,newtxmath}

\usepackage[T1]{fontenc}

\DeclareRobustCommand{\VAN}[3]{#2}
\let\VANthebibliography\thebibliography
\def\thebibliography{\DeclareRobustCommand{\VAN}[3]{##3}\VANthebibliography}

\usepackage{graphicx}	
\usepackage{amsmath}	

\title[COMs in Turbulent Disks]{Chemical Evolution of Complex Organic Molecules in Turbulent Protoplanetary Disks: Effect of stochastic UV irradiation}

\author[T. Suzuki et al.]{
T. Suzuki,$^{1}$\thanks{E-mail:taikisuzuki@g.ecc.u-tokyo.ac.jp}
K. Furuya,$^{2}$
Y. Aikawa,$^{1}$
T. Shibata,$^{2}$
and L.Majumdar$^{3,4}$
\\
$^{1}$Department of Astronomy, The University of Tokyo, Bunkyo-ku, Tokyo 113-0033, Japan\\
$^{2}$National Astronomical Observatory of Japan, Osawa 2-21-1, Mitaka, Tokyo 181-8588, Japan\\
$^{3}$School of Earth and Planetary Sciences, National Institute of Science Education and Research, Jatni 752050, Odisha, India\\
$^{4}$Homi Bhabha National Institute, Training School Complex, Anushaktinagar, Mumbai 400094, India\\
}

\date{Accepted XXX. Received YYY; in original form ZZZ}

\pubyear{2024}
\begin{document}
\label{firstpage}
\pagerange{\pageref{firstpage}--\pageref{lastpage}}
\maketitle

\begin{abstract}
We investigate the chemical evolution of complex organic molecules (COMs) in turbulent disks using gas-ice chemical reaction network simulations. We trace trajectories of dust particles considering advection, turbulent diffusion, gas drag, and vertical settling, for 10$^6$ yrs in a protoplanetary disk. Then, we solve a gas-ice chemical reaction network along the trajectories and obtain the temporal evolution of molecular abundances. 
We find that the COM abundances in particles can differ by more than two orders of magnitude even when the UV fluence (i.e., the time integral of UV flux) received by the particles are similar, suggesting that not only the UV fluence but also the time variation of the UV flux does matter for the evolution of COMs in disks.
The impact of UV fluence on molecular abundances differs between oxygen-bearing and nitrogen-bearing COMs. 
While higher UV fluence results in oxygen being locked into CO$_2$, leading to reduced abundances of oxygen-bearing COMs such as CH$_3$OCH$_3$, mild UV exposure can promote their formation by supplying the precursor radicals.
On the other hand, nitrogen is not locked up into specific molecules, allowing the formation of nitrogen-bearing COMs, particularly CH$_3$NH$_2$, even for the particle that receives the higher UV fluence.
We also find that the final COM abundances are mostly determined by the inherited abundances from the protostellar core when the UV fluence received by dust particles is less than a critical value, while they are set by both the inherited abundances and the chemistry inside the disk at higher UV fluence.
\end{abstract}




\section{Introduction} \label{sec:intro}
Complex organic molecules (COMs), which contain six atoms or more, around low-mass young stellar objects (YSOs) have been a topic of great interest, with many observations carried out using radio telescopes towards the inner, hot ($\gtrsim$100 K) regions of envelope and circumstellar disks.
The protostellar sources with the detection of COMs such as methanol (CH$_3$OH), methyl formate (HCOOCH$_3$), and dimethyl ether (CH$_3$OCH$_3$) are termed as hot corinos \citep[e.g.,][]{Dishoeck95,Schoier02,Cazaux03,Ceccarelli06}.
The ALMA-PILS survey, for example, revealed a diverse chemical inventory of COMs in the envelope around Class 0 protostellar binary, IRAS~16293–2422 \citep{Jorgensen16,Manigand20}.
Such COMs could be passed on to the protoplanetary disks, potentially laying the ground for planetary material rich in organic molecules. \citet{drozodvskaya19}, for example, found a positive correlation between the COM abundances in the envelope gas around IRAS~16293-2422 and those in the coma of Solar system comet 67P/Churyumov-Gerasimenko.

The detection of COMs in protoplanetary disks was historically challenging due to the cold nature of the bulk of disks.
In typical disks around T Tuari stars, the midplane temperature outside several au is below 100 K, causing COMs, if exist, to freeze out onto grains.
Until very recently, the detection of COMs was limited to CH$_3$OH and CH$_3$CN, whose emission is weak and spatially extended beyond their snowlines. They could be non-thermally desorbed from ices or formed in the gas-phase from the sublimates \citep{Oberg15,Walsh16,Loomis18,Bergner18}.
%
%
%
%
Observations of peculiar warm disks recently opened a new window to the observations of COMs in disks.
In the disk of FU Ori object V883~Ori, various icy COMs are sublimated due to a temporal luminosity outburst. \citet{Lee19} found that the abundances of CH$_3$COCH$_3$, CH$_3$CHO, and CH$_3$OCHO in the V883~Ori disk exceed those in IRAS 16293B by factors ranging from 4 to 16 \citep[see also][]{Yamato24}.
\cite{Brunken22} observed a transitional disk around Oph~IRS~48, which is a Herbig A0 star, and detected CH$_3$OCH$_3$ and CH$_3$OCHO at the dust trap near the edge of the central cavity.
With a higher angular resolution of 0.3", \cite{Booth24} report that the abundances of CH$_3$OCHO and CH$_3$OCH$_3$ relative to CH$_3$OH are 0.28 ± 0.04 and 0.25 ± 0.03, respectively, which are higher than those observed in hot corinos.
These observations indicate that COMs (other than CH$_3$OH) might have efficiently formed inside the disks.

On the other hand, some disk observations have pointed out that the chemical composition might be inherited from the protostar era.
\citet{Booth21a} reported the detection of CH$_3$OH in the disk around a Herbig Be star HD 100546, where the formation of CH$_3$OH inside the disk would be hindered due to the warm temperatures ($>$20 K).
Then CH$_3$OH detected in the HD 100546 disk is likely inherited from the earlier evolutionary stage \citep{Booth21a}.
\cite{Brunken22} pointed out that the observed CH$_3$OCH$_3$/CH$_3$OCHO ratio is consistent across diverse environments of hot corinos, disks, and comets, suggesting a chemical connection between the two species.
\cite{Tobin23} found that the CH$_2$DOH/CH$_3$OH ratio is approximately ten times the HDO/H$_2$O ratio in all of the coma of comet 67P/Churyumov-Gerasimenko, protostars, and the protoplanetary disk of V883 Ori. They highlighted this as evidence that these molecules, formed during the prestellar phase, are carried over to comets.
It is becoming increasingly clear that both inheritance from the protostellar core and in-disk chemical processes play roles in the presence and abundances of COMs in protoplanetary disks and ultimately in the planetary systems.

Chemical network simulation is a powerful tool to study if and how COMs are formed and destroyed in protoplanetary disks. It is well established that the disks have layered structures with different molecular compositions in the vertical direction \citep[e.g.,][]{bergin07}.
%
%
Radicals are abundant in the disk surface layer due to the strong stellar UV radiation, while COMs are abundant as ice in the cold midplane outside the water snow line  \citep[e.g.,][]{Aikawa02,Walsh14,Ruaud19}.
%
%
%
%
Turbulent mixing can affect the abundance and spatial distribution of COMs in disks. 
\citet{Semenov11} and \citet{Furuya14}, for example, performed the chemical reaction network simulations in disks with turbulent mixing in the Eulerian manner, assuming that the gas and grains are well coupled.
%
%
%
%
%
%
%
%
They showed that stable molecules formed on grain surfaces in the midplane are transported to the disk surface to be sublimated and/or photodissociated, while radicals formed in the disk surface are transported to the deeper layers to activate chemical reactions including the formation of COMs.
%
It is worth noting that in the Eulerian approach, only the mean compositions of gas and ice are obtained at each spatial grid in the disk. The ice composition, however, could vary among grains in the same spatial grid at a given time; the motion of grain is random in turbulent gases and grains have different thermal and UV-irradiation histories.

%
%
%
%
%
%

The dust tracking model with the Lagrangian approach \citep{Ciesla10,Ciesla11} is an alternative and complementary approach to understanding the chemical evolution in ice mantles of dust grains.
\cite{Ciesla12} used this approach to calculate the cumulative number of UV photons irradiated on dust grains. They argued that ample organic molecules could form, {\it assuming} that the abundances of COMs are proportional to the cumulative number of UV photons.
%
%
%
\cite{Bergner21} also employed the dust-tracking approach to study the survival of icy molecules in disks. By focusing solely on the destruction of ices, i.e. desorption and dissociation by UV photons, they showed that the inheritance of interstellar ices to comets requires pebble formation outside a few tens of au, since ice destruction is rapid for small grains ($< 10$ $\mu$m) in the inner radius.
%
%
%
\cite{Takehara22}, on the other hand, applied a graph-theoretic matrix framework to study the rearrangement of chemical bonds in the ice mantle of dust grains in the presence of UV photons. When the dust grains are small, they are stirred up to the disk surface, where various unstable molecules are formed by photodissociation. 
As the grains grow by coagulation and sediment to the disk midplane, photodissociation reactions are quenched, and sugars and other stable complex organic molecules are formed by recombination of radicals in their model.
While the application of graph theorem enabled them to investigate the formation of large molecules such as sugar, the results should be taken with caution, since they neglect the details of ice chemistry, e.g., activation barriers of diffusion and reactions and desorption of molecules to the gas phase.
%
%

%
%
In the present work, we perform chemical network calculations of gas-grain chemistry based on the rate equation approach, including both destruction and production of molecules, combined with the Lagrangian dust-tracking approach to investigate the formation and destruction of COMs in protoplanetary disks.
The rest of the paper is organized as follows.
In Section 2, we describe our trajectory calculations and physical and chemical models.
In Section 3, we discuss how the stochastic UV irradiation on dust grains due to the turbulent diffusion affects the abundances of major element carriers and COMs.
In Section 4, we analyze the variation of COM abundances in individual UV exposure events and discuss the dependence of our results on the parameters in the chemical model.
Our findings are summarized in Section 5.

\section{Method}
By tracing the trajectories of grain particles in a disk model, we obtain the temporal variation of physical parameters such as UV flux, temperature, and gas density that each particle experiences for 10$^{6}$ years.
Under those physical parameters, chemical network calculations are performed to investigate the evolution of molecular abundances in the gas and ice mantles.

\subsection{Physical Model}
\begin{figure*}[ht!]
    \begin{tabular}{cc}
      \begin{minipage}[t]{0.45\hsize}
        \centering
        \includegraphics[keepaspectratio, scale=0.5]{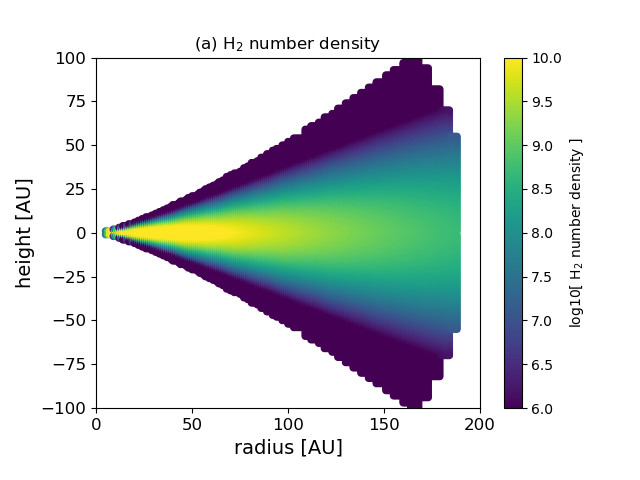}
        \label{composite}
      \end{minipage} &
      \begin{minipage}[t]{0.45\hsize}
        \centering
        \includegraphics[keepaspectratio, scale=0.5]{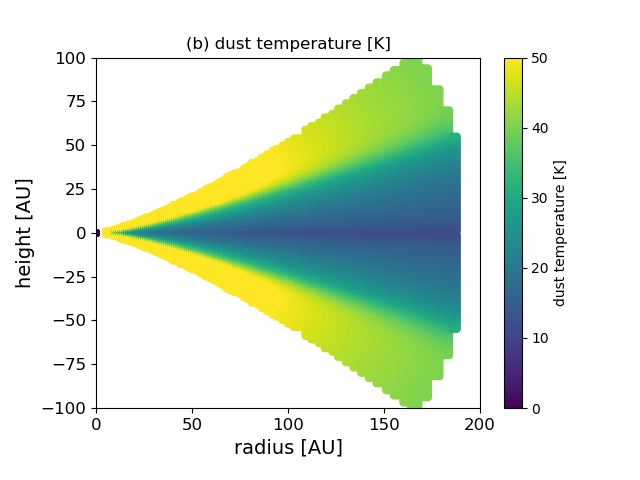}
        \label{Gradation}
      \end{minipage} \\
   
      \begin{minipage}[t]{0.45\hsize}
        \centering
        \includegraphics[keepaspectratio, scale=0.5]{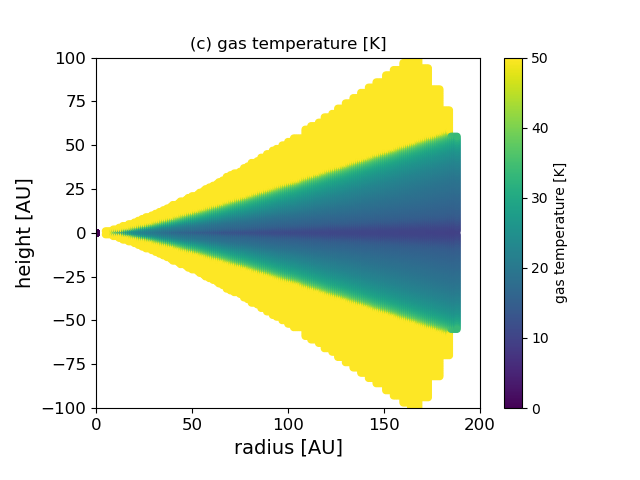}
        \label{fill}
      \end{minipage} &
      \begin{minipage}[t]{0.45\hsize}
        \centering
        \includegraphics[keepaspectratio, scale=0.5]{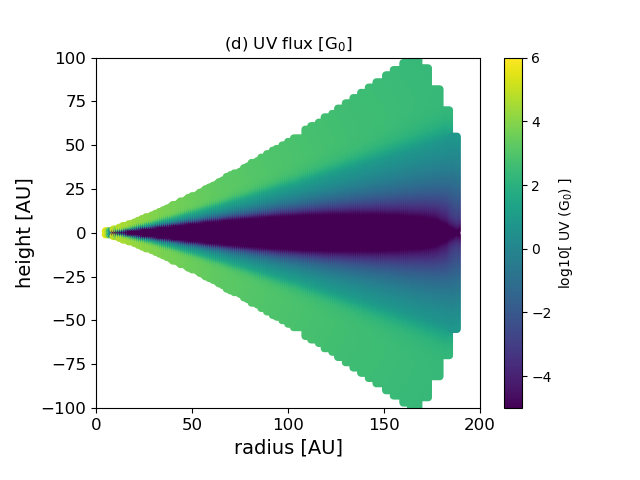}
        \label{transform}
      \end{minipage} 
    \end{tabular}
    \caption{The spatial distributions of (a) gas density, (b) dust temperature, (c) gas temperature, and (d) wavelength-integrated UV flux in the disk model. The integrated UV flux is normalized by the interstellar UV flux and given in a unit of $G_0$ \citep{draine78}.}
    \label{fig:physical_model}
\end{figure*}

We adopt a two-dimensional physical model of a Class II disk around TW~Hya, assuming axisymmetry \citep{Furuya22}.
The gas and dust density distributions are taken from \cite{Cleeves15}.
Dust temperature and UV flux at each point in the disk are obtained by solving radiative transfer with RADMC-3D \citep{Dullemond12}.
The input UV spectra of TW Hya are taken from \citet{dionatos19}, and they are added to a blackbody
component with an effective temperature of 4110 K \citep{andrews12}.
The gas temperature is obtained by calculating the heating and cooling together with a chemical network to derive the abundances of coolant species at each position in the disk.
The cosmic-ray ionization rate is set to be constant, 1.3$\times$10$^{-17}$ s$^{-1}$.
Figure~\ref{fig:physical_model} shows the spatial distributions of (a) gas density, (b) dust temperature, (c) gas temperature, and (d) the sum of the stellar and interstellar FUV flux integrated from 912 to 2000  \AA  in our disk model.
When addressing the UV flux, we adopt the interstellar Draine field $\text{G}_0$ as a normalization factor.  It is equivalent to \( 2 \times 10^8 \, \text{photon cm}^{-2}\,\text{s}^{-1} \).
We note that the cosmic-ray (CR) induced UV radiation \citep[e.g.,][]{Prasad83,Gredel87} is considered in the chemical reaction network (see below), while it is not included in Figure~\ref{fig:physical_model} (d).
In addition to UV, protoplanetary disks are irradiated by stellar X-rays, which is an important ionization source of the disks.
Throughout this paper, we neglect stellar X-rays to isolate the impact of stochastic UV radiation on chemistry.

\begin{figure}
 \begin{tabular}{l}
\includegraphics[scale=0.6]{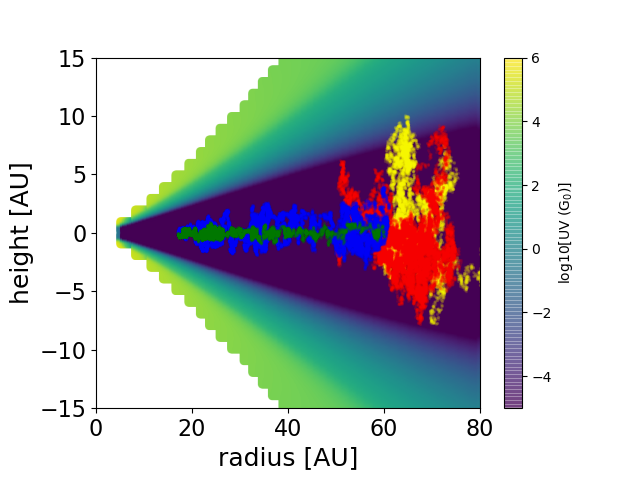}
 \end{tabular}
\vspace{15mm}
\caption{
The example of trajectories of particles which are initially located in the midplane at 80 au from the central star. The trajectories were tracked for 10$^6$ yr.
The trajectories of the particles with the size of 1, 10, 100, and 1000~$\micron$ are shown in yellow, red, blue, and green, respectively.
\label{fig:orbits_overUV}
}
\end{figure}

We follow the numerical method developed in \cite{Ciesla10} and \cite{Ciesla11} to obtain the trajectory of each grain particle in the disk. 
This method incorporates the processes of turbulent diffusion, advection, gravitational settling, and gas drag to accurately depict the motion of a grain within the disk. 
The local diffusivity is given by $\alpha c_s(R,z) H(R)$, where $\alpha = 10^{-3}$, $c_s(R,z)$ is the sound speed, and $H(R)$ is the gas scale-height, the latter two of which are taken from our disk physical model.
Although our disk model is two-dimensional, a three-dimensional Cartesian coordinate system is used to track the motion of each grain, assuming axisymmetry.
%
%
%
The position and velocity at a given time step are recorded and used to calculate the next time step.
We set the size of a time step, $\Delta t$, to be 1/25 of the local orbital period.
Note that in the particle tracking method, either the growth or fragmentation of the grains through collisions is not considered.
When a particle enters inside 1~au from the central star during the calculation, we assumed that the particle fell into the central star and terminated the calculation.
As a benchmark for our calculation code, we confirmed that our code reproduces the trajectories of particles of different sizes shown in Figure~1 of \cite{Bergner21}.
%
%
Figure~\ref{fig:orbits_overUV} depicts the example of the trajectories in 10$^6$ yr for 1, 10, 100, and 1000~$\micron$ sized grains starting from $R = 80$ au and $z = 0$ au.
The color scale in the background shows the distribution of FUV flux (Figure \ref{fig:physical_model}d).
Smaller grains tend to be lifted up to higher $z$ by turbulence.
Larger particles tend to remain near the midplane, and fall towards the central star by radial drift.
%

%

%

Since the UV exposure experienced by particles varies over time due to turbulent motion, it is useful to define the total amount of UV radiation a particle receives during its trajectory; ``fluence'' is the time integral of UV flux with a unit of cm$^{-2}$ in cgs.
In addition to the UV radiation from the central star and the interstellar radiation field, we also consider the CR-induced UV.
CRs can penetrate even in dense regions where the external UV is heavily attenuated, producing UV radiation through the excitation of H$_2$.
%
%
Assuming the CR ionization rate of 1.3$\times$10$^{-17}$~s$^{-1}$,
the flux of CR-induced UV is assumed to be $1.5 \times 10^{-5} \, \text{G}_0$ everywhere in our disk model \citep[corresponds to $3\times10^3$ photons cm$^{-2}$ s$^{-1}$;][]{shen04}.
\footnote{CRs could be slightly attenuated near the midplane inside the disk radius of several au. We however assume a constant ionization rate in our disk model for simplicity, since we aim to investigate the correlation between COMs abundance and cumulative UV flux rather than to calculate their spatial distribution.}
As the CR-induced UV determines the lower limit of the UV flux, it also sets the lower bound for the fluence experienced by a particle, and the total UV fluence of CR-induced UV over 10$^6$ yr is $4.5\times10^{8} \ \text{Yr} \text{G}_0$.
\footnote{Strictly speaking, the reactions caused by the comic-ray-induced UV photons are slightly different from those caused by stellar and interstellar UV photons. The reaction rates of the former reactions are calculated separately from the latter in our network model. It is, however, helpful to use the CR-induced UV photons as the normalization of the fluence, since their fluence is basically constant for all grain particles, unless they fall to the central star before $10^6$ yr.}
We define the normalized fluence $\Gamma$ as:
\begin{equation}
\Gamma = \frac{\int_0^{t_{\rm max}} [F_{\ast}(\vec{l}(t)) + F_{\rm ISUV}(\vec{l}(t)) + F_{\rm CRUV}] dt}{\int_0^{10^6 {\rm yr}} F_{\rm CRUV} dt}, \label{eq:ft}
\end{equation}
where $F_{\ast}$, $F_{\rm ISUV}$, and $F_{\rm CRUV}$ are the local flux of the stellar, interstellar, and CR-induced UV photons, respectively, $t_{\rm max}$ is $10^6$ yr or the time for the grain to reach the $r=1$ au boundary, and $\vec{l}(t)$ is the position vector of a particle at a given time $t$ \citep{Furuya17}.
%
%
Since the CR-induced UV photons set the minimum of the UV radiation field in our model, 
the normalized fluence can be used as a measure of the importance of the stellar and interstellar UV radiation.

\begin{figure}
 \begin{tabular}{l}
\includegraphics[scale=0.6]{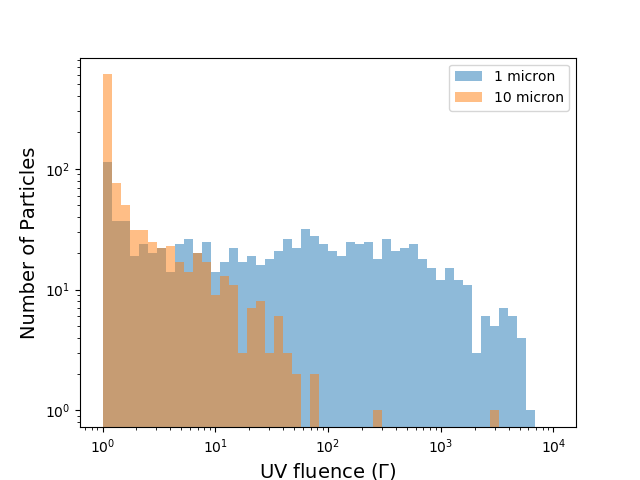}
 \end{tabular}
\vspace{15mm}
\caption{
The frequency distribution of $\Gamma$, which is defined by Equation \ref{eq:ft}. The blue and red colors show the cases for a particle size of 1 and 10 $\micron$, respectively.
\label{fig:histogram_uv}
}
\end{figure}

We calculate the trajectories of 1000~particles each for grains of 1 $\micron$ and 10 $\micron$ for $10^6$ yrs.
The initial position of particles is randomly assigned but follows the gas density distribution of the disk physical model (Figure~\ref{fig:physical_model} a).
Figure~\ref{fig:histogram_uv} shows the frequency distribution of the FUV fluence for the 1 $\micron$ and 10 $\micron$ sized grains.
The smaller grains tend to be stirred up to the disk surface by turbulence and thus obtain larger fluence.
In this study, we focus on small dust particles to examine the impact of UV radiation on molecular evolution. We use 1~$\micron$ dust particles for the main discussion, and the results for 10~$\micron$ particles are provided in the Appendix.

\subsection{Chemical Network Simulation} \label{sec:network}
We utilize the gas-ice astrochemical code based on the rate equation approach 
 \citep[Rokko code;][]{Furuya15}. The gas-ice chemistry is described by the three-phase model \citep{hasegawa93}, taking into account gas-phase chemistry, interactions between gas and (icy) grain surfaces, grain surface chemistry, and chemistry inside the bulk ice mantle.
The gas-ice chemical network is taken from \cite{Garrod13}.
It includes COMs formation both in the gas phase and on grains. In the gas phase, molecules are photodissociated by stellar and interstellar UV radiation and cosmic-ray induced UV. Molecules interact through neutral-neutral reactions such as HCO + CH\(_2\)OH $\rightarrow$ CH\(_3\)OH + CO, and ion-neutral reactions followed by dissociative recombination with electrons, for example, CH\(_3\)OH\(_2^+\) + e\(^-\) $\rightarrow$ CH\(_3\)OH + H. Our model does not include gas phase formation processes for larger COMs than CH$_3$OH. However, since dust temperatures within  our model are mostly below 40~K, inclusion of gas-phase formation of large COMs would not significantly change our results.
On the grains, radicals are produced via photodissociation. The network includes radical-radical reactions such as CH\(_3\) + OH $\rightarrow$ CH\(_3\)OH and hydrogen abstraction reactions such as CH\(_3\)O + HNCO $\rightarrow$ CH\(_3\)OH + OCN, whose efficiency is governed by the thermal hopping rate of the reactants and the activation barriers of the reactions. The radical-radical reactions on grain surfaces, which are crucial for the formation of COMs in our models, are assumed to be barrier-less.
We note that this assumption may results in overestimation of some COMs (e.g. see \S 3.1).

The top four monolayers (MLs) are considered as surfaces, refering to \cite{Vasyunin13}. They showed that considering top four layers instead of a single layer better reproduces the results of temperature-programmed desorption experiments for CO and CO$_2$ mixed with H$_2$O ice.
The rest is considered as the bulk ice mantle.
While the bulk ice mantle was assumed to be chemically inert in \citet{Furuya15}, we assume that photodissociation and two-body reactions occur in the bulk ice mantle, following the method in \citet{Garrod13}.
However, we do not consider the swapping between surface species and mantle species to save the computational cost.

%
%
%
We calculate the photodissociation rates by stellar and interstellar UV photons in the gas phase by scaling the rates in \citet{Garrod13} (which assumes the interstellar UV radiation field), referring to the wavelength-integrated stellar and interstellar FUV flux at each point in the disk \citep{Walsh10}.
The self-shielding and mutual shielding factors for the photodissociation of H$_2$, CO, and N$_2$ are taken from \citet{draine96}, \citet{visser09}, and \citet{li13}, respectively.
The molecular column densities for calculating the shielding factors are taken from the physical and chemical model of the TW Hya disk \citep{Furuya22}.
Note that the attenuation of UV radiation by dust is already accounted for in the calculation of the UV radiation field in the disk.
%
%
%
%
The photodissociation rates of species $i$ ( cm$^{-3}$~s$^{-1}$ ) on the grain surface and in the ice mantles are calculated in a similar way to \citet{Furuya15}:
\begin{align}
R^{(s)}_{\rm phdiss,\,i} &= n_{\text{gr}}F_{\rm UV} \sigma_{\rm gr} f^{(s)}_i  P_{\text{abs,} i} \times 4, \label{eq:phdiss_s} \\
R^{(m)}_{\rm phdiss,\,i} &= n_{\text{gr}}F_{\rm UV} \sigma_{\rm gr} f^{(m)}_i  P_{\text{abs,} i}\times 
\Sigma_{j=1}^{N_{\rm layer}}(1-f^{(m)}_i P_{\text{abs,} i})^{j-1}, \label{eq:phdiss_m}
\end{align}
where \( n_{\text{gr}} \) denotes the number density of dust grains per unit volume, $F_{\rm UV}$ is the sum of the stellar FUV radiation, interstellar ultraviolet radiation, and CR-induced UV, $\sigma_{\rm gr}$ is the geometrical cross-section of a dust grain, $f^{(s)}_i$ and $f^{(m)}_i$ are the fractional abundance of species $i$ w.r.t all ice species on the surface and in the bulk ice mantle, respectively, and $P_{\rm{abs,} i}$ is the absorption probability of the incident FUV photon per monolayer of species $i$.
$N_{\rm layer}$ is the number of monolayers in the bulk ice mantle.
The factor of four in Eq. \ref{eq:phdiss_s} refers to the absorption by the four outermost layers (i.e., the surface) \citep{Vasyunin13}.
Eq. \ref{eq:phdiss_m} accounts for the attenuation of UV photons inside ice mantles, assuming that species $i$ in upper ice layers shields species $i$ in lower ice layers from UV, but does not shield other species in lower ice layers (i.e., the overlap of photoabsorption bands between species $i$ and other species is assumed to be negligible). 
As a result, the attenuation is important only for the most abundant molecules such as H$_2$O and CO$_2$.
We discuss the impact of this assumption on our results in Section \ref{sec:photodiss}.

The diffusion-to-desorption activation energy ratio ($\chi$) is the key parameter for two-body reactions on grain surfaces and in bulk ice mantles, although it is poorly quantified and the value is most likely species-dependent \citep{Furuya22diff}.
In our fiducial model, $\chi$ is set to 0.4 for the surface chemistry and 0.8 for the bulk ice mantle chemistry.
We discuss the impact of the $\chi$ value for the bulk ice mantle chemistry on our results in Section \ref{sec:ebed_ratio}.

The CR ionization rate of H$_2$ is set to be 1.3$\times$10$^{-17}$ s$^{-1}$ throughout our calculations.
Our model includes all the cosmic-ray-related chemical processes both in the gas phase and on grain surfaces, including the induced UV and non-thermal desorption via the stochastic heating of grains.

Initially, we assume that the composition of ice mantles is the same for all dust particles, and C, N, O, and S are fully locked into ice molecules.
Their abundances are motivated by the observation of interstellar ices \citep{Boogert15}.
The abundances of other heavy elements are taken from \citet{aikawa99}, and are assumed to be in the form of either neutral atoms or atomic ions, depending on their ionization energy, while H is all in H$_2$.
The initial abundances are listed in Table~\ref{tbl:initial_abundance}.
\begin{table}
\centering
\caption{Initial Abundances in the Fiducial and Inherited Models}
\label{tbl:initial_abundance}
\begin{tabular}{llll}
\hline
Species & Abundance (/H) & Species* & Abundance* (/H) \\
\hline
H$_2$O ice & 1.0$\times$10$^{-4}$ & CH$_3$OCH$_3$ ice$^*$ & 2.0$\times$10$^{-7}$\\
CO ice & 5.0$\times$10$^{-5}$ & HCOOCH$_3$ ice$^*$ & 1.0$\times$10$^{-7}$\\
CO$_2$ ice & 3.0$\times$10$^{-5}$ & CH$_3$COCH$_3$ ice$^*$ & 9.2$\times$10$^{-9}$\\
CH$_4$ ice & 5.0$\times$10$^{-6}$ & C$_2$H$_5$OH ice$^*$ & 3.1$\times$10$^{-8}$\\
CH$_3$OH ice & 5.0$\times$10$^{-6}$ & NH$_2$CHO ice$^*$ & 7.7$\times$10$^{-10}$\\
N$_2$ ice & 7.4$\times$10$^{-6}$ & & \\
NH$_3$ ice & 1.0$\times$10$^{-5}$ & & \\
H$_2$S ice & 9.2$\times$10$^{-8}$ & & \\
Si$^+$ & 9.7$\times$10$^{-9}$ & & \\
Fe$^+$ & 2.7$\times$10$^{-9}$ & & \\
Na$^+$ & 2.3$\times$10$^{-9}$ & & \\
Mg$^+$ & 1.1$\times$10$^{-8}$ & & \\
Cl$^+$ & 2.2$\times$10$^{-10}$ & & \\
P$^+$ & 1.0$\times$10$^{-9}$ & & \\
\hline
\end{tabular}
\smallskip\\
{\footnotesize Species marked with a '*' are only included in the initial conditions of the Inherited Model (see Section \ref{sec:inherit}).}
\end{table}

In the chemical network simulations, we always set the gas-to-dust mass ratio to 100, and the total (gas+ice) elemental abundances are conserved in each particle.
In other words, we assume that the gas follows the motion of a dust particle, and there is no exchange of materials between the background disk gas and our closed parcel of gas and dust.
%
%
Nonetheless, because the range of temperatures experienced by the particles in our simulations is typically from 20 K to 40 K, meaning that most molecules remain frozen onto the dust grains, this simplification would not significantly affect the results, in particular, the formation of COMs via radical-radical reactions in ice mantles.

\section{Results}
\subsection{Temporal Variation of Molecular Abundances in an Example Particle} \label{sec:abundances_example}
\begin{figure*}
 \centering
\begin{minipage}[t]{0.3\textwidth}
 \includegraphics[width=5.5cm]{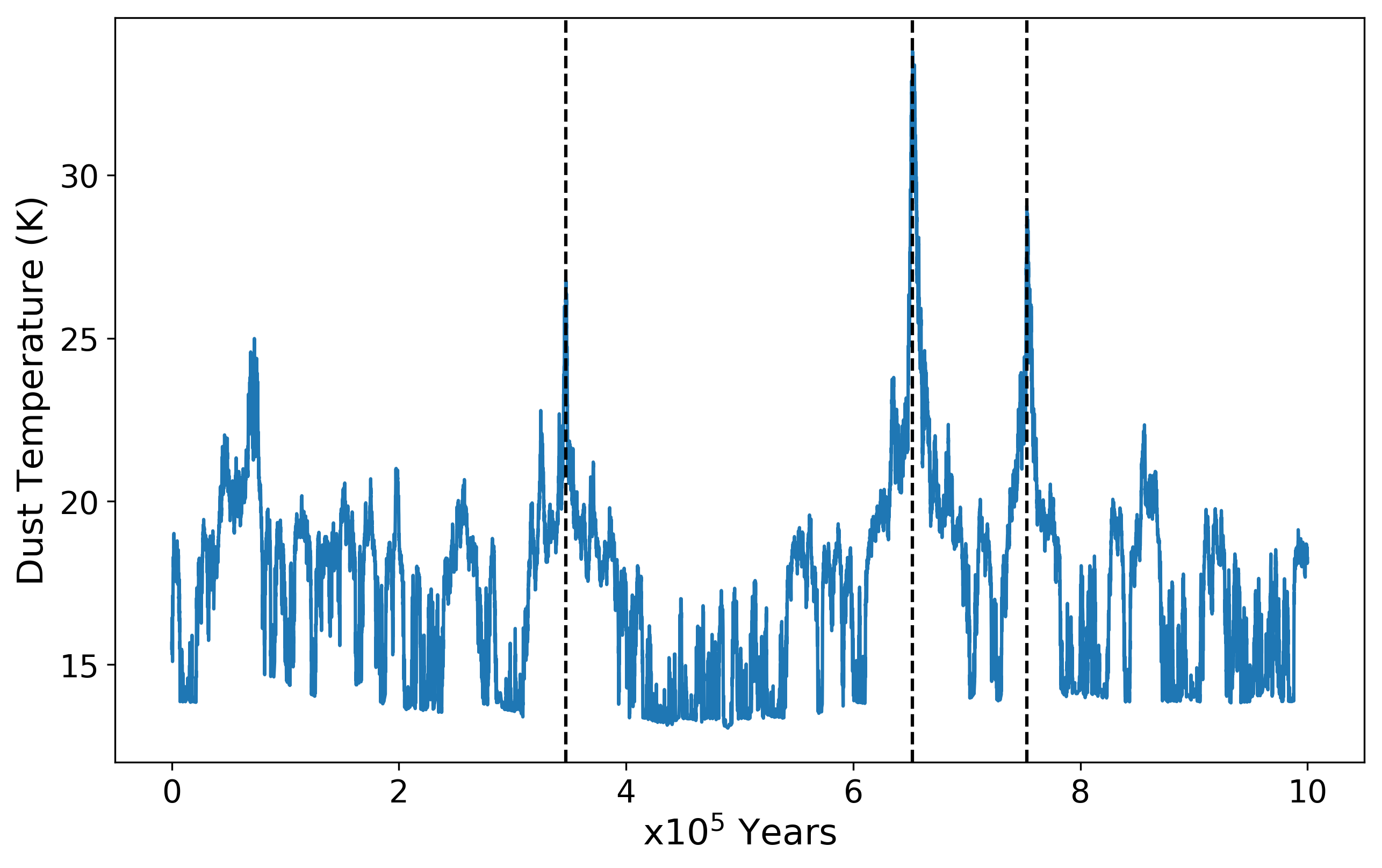}
\end{minipage}
\begin{minipage}[t]{0.3\textwidth}
 \includegraphics[width=5.5cm]{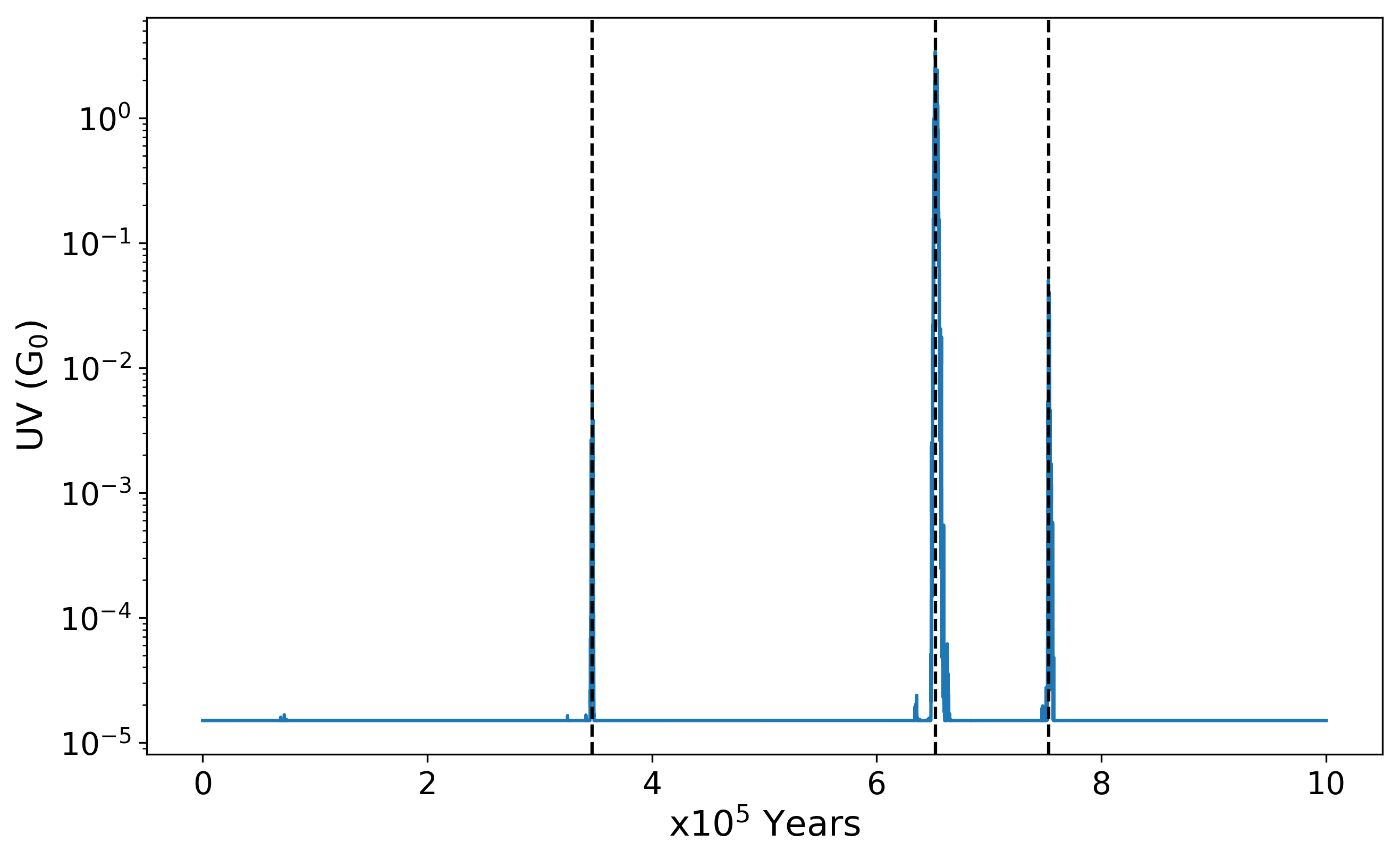}
\end{minipage}
\begin{minipage}[t]{0.3\textwidth}
 \includegraphics[width=5.5cm]{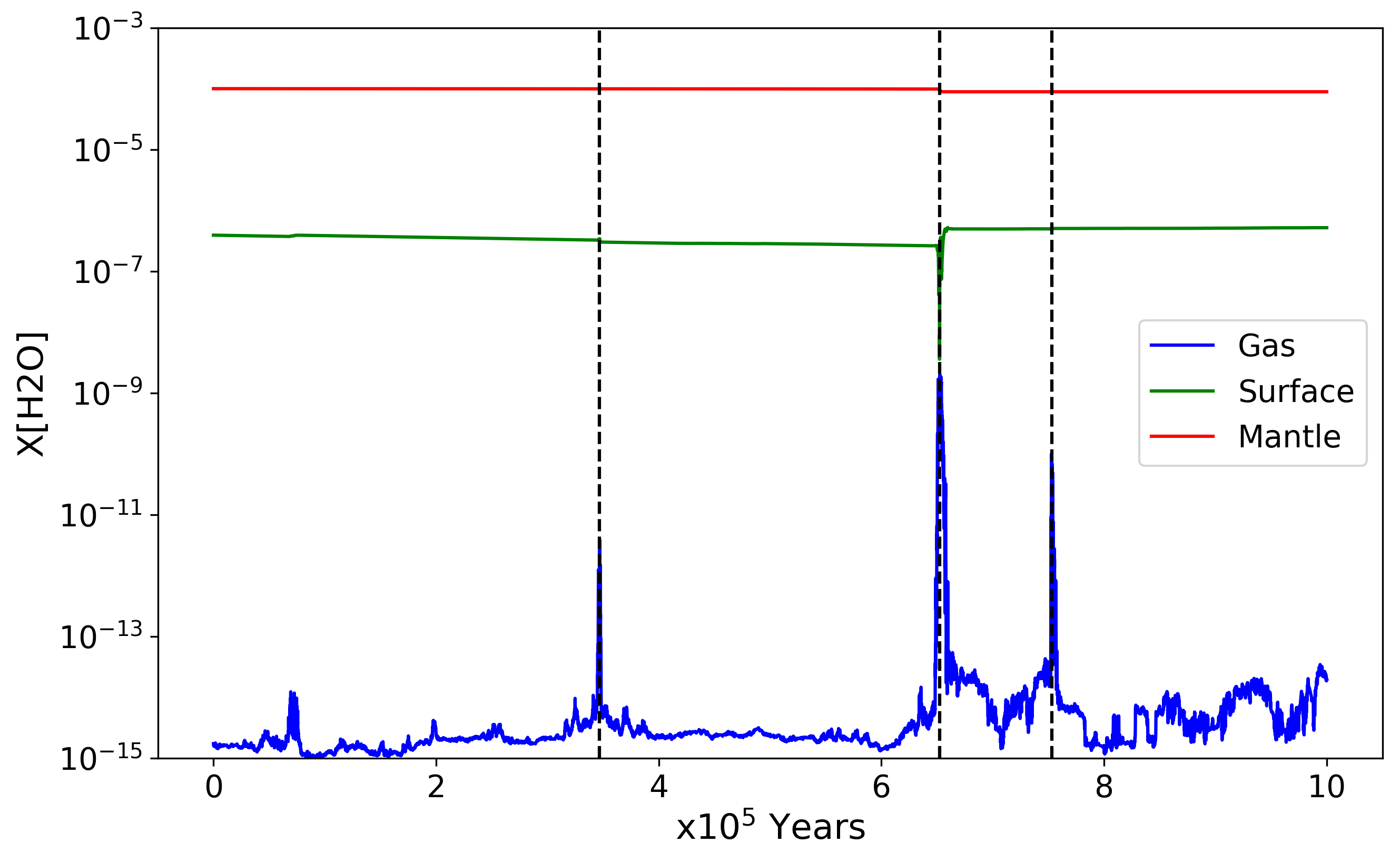}
\end{minipage}
\begin{minipage}[t]{0.3\textwidth}
 \includegraphics[width=5.5cm]{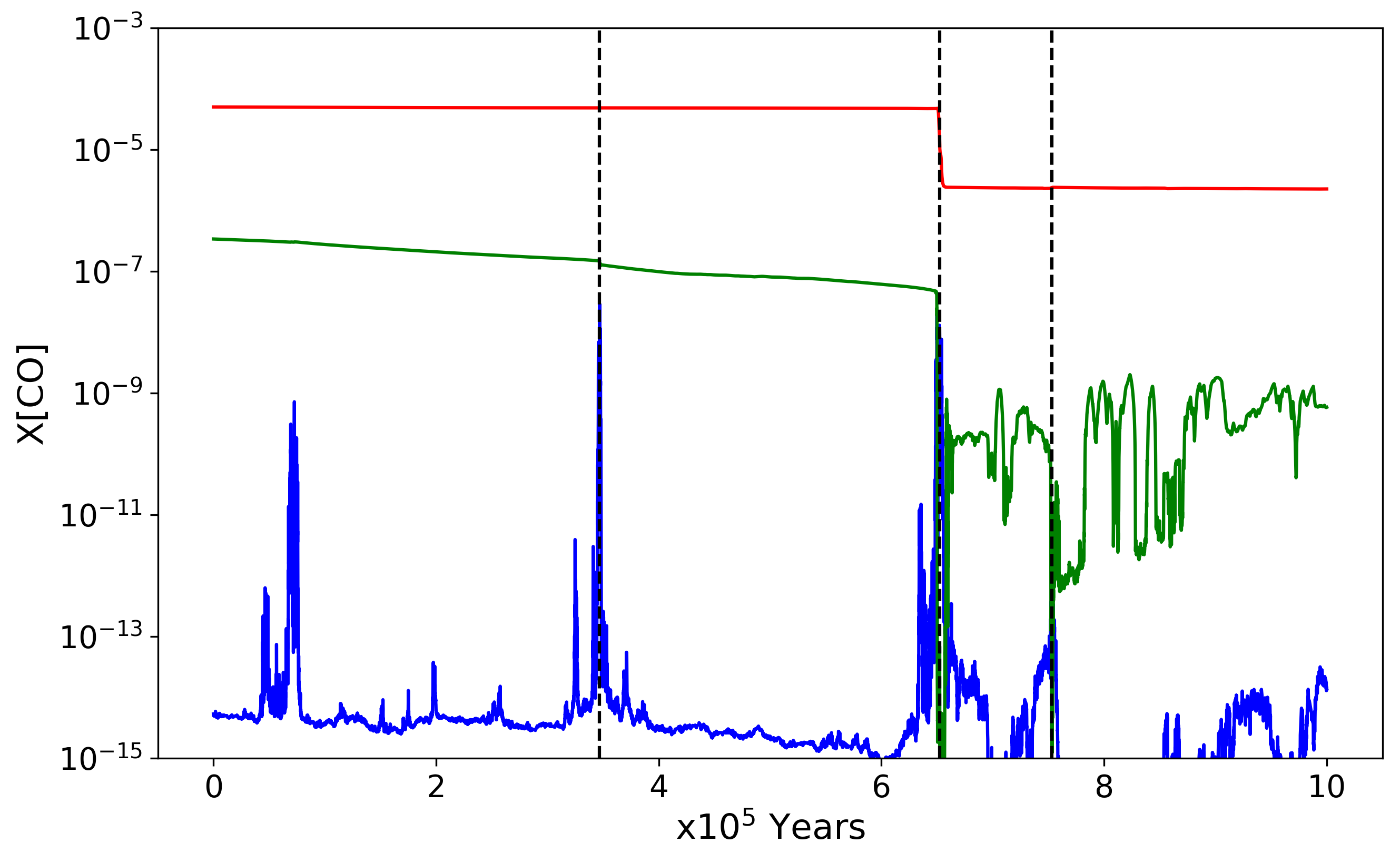}
\end{minipage}
\begin{minipage}[t]{0.3\textwidth}
 \includegraphics[width=5.5cm]{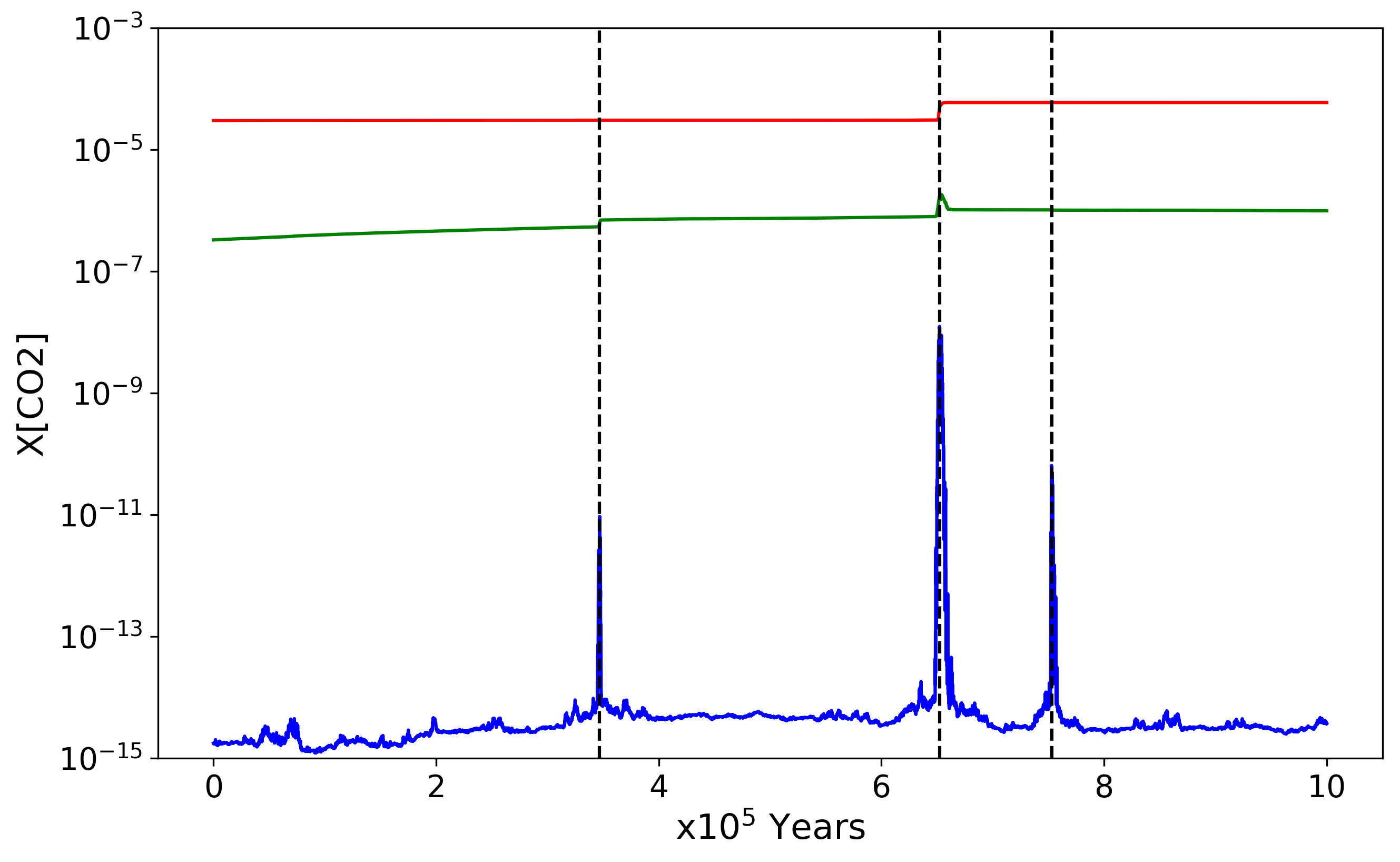}
\end{minipage}
\begin{minipage}[t]{0.3\textwidth}
 \includegraphics[width=5.5cm]{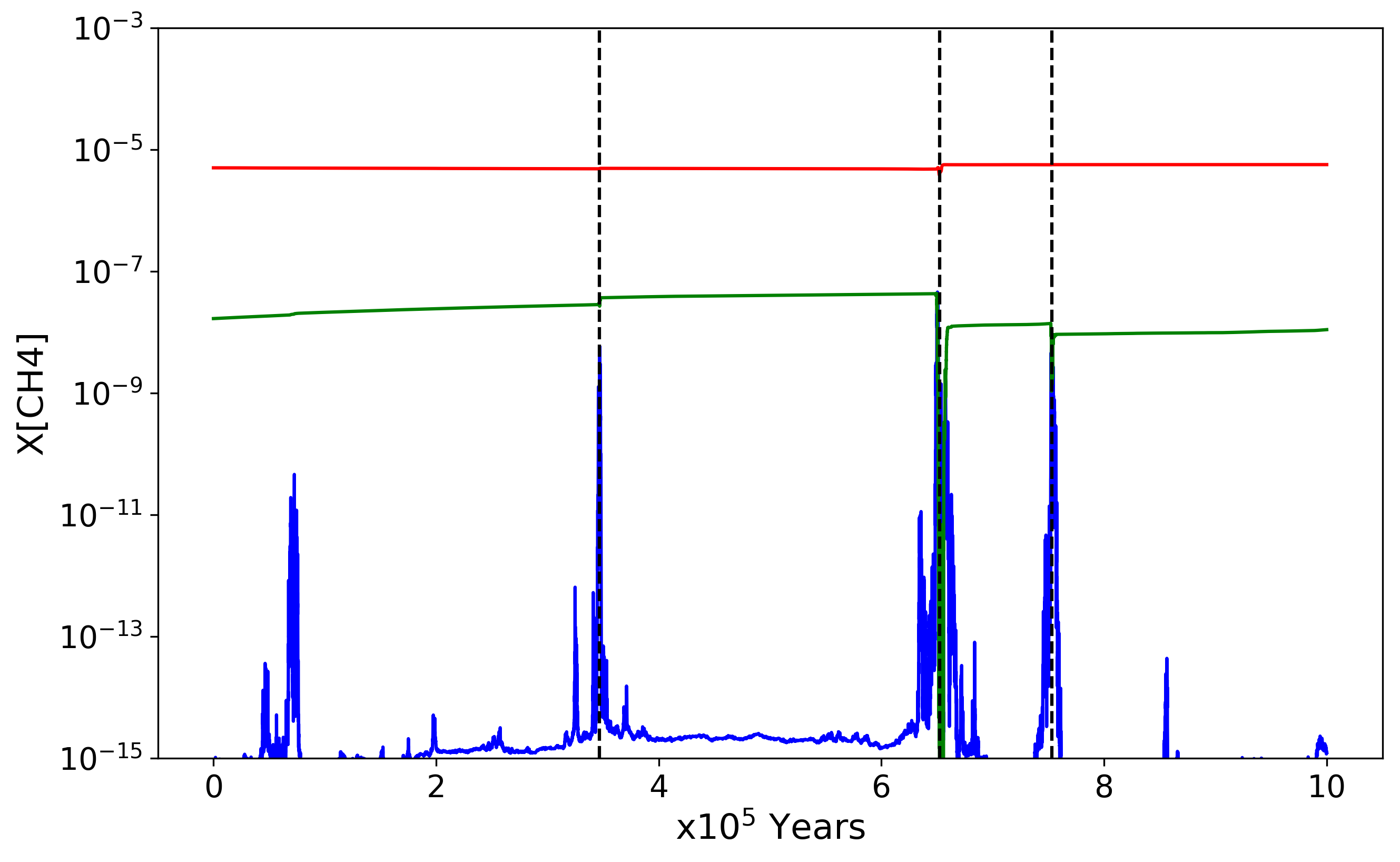}
\end{minipage}
\begin{minipage}[t]{0.3\textwidth}
 \includegraphics[width=5.5cm]{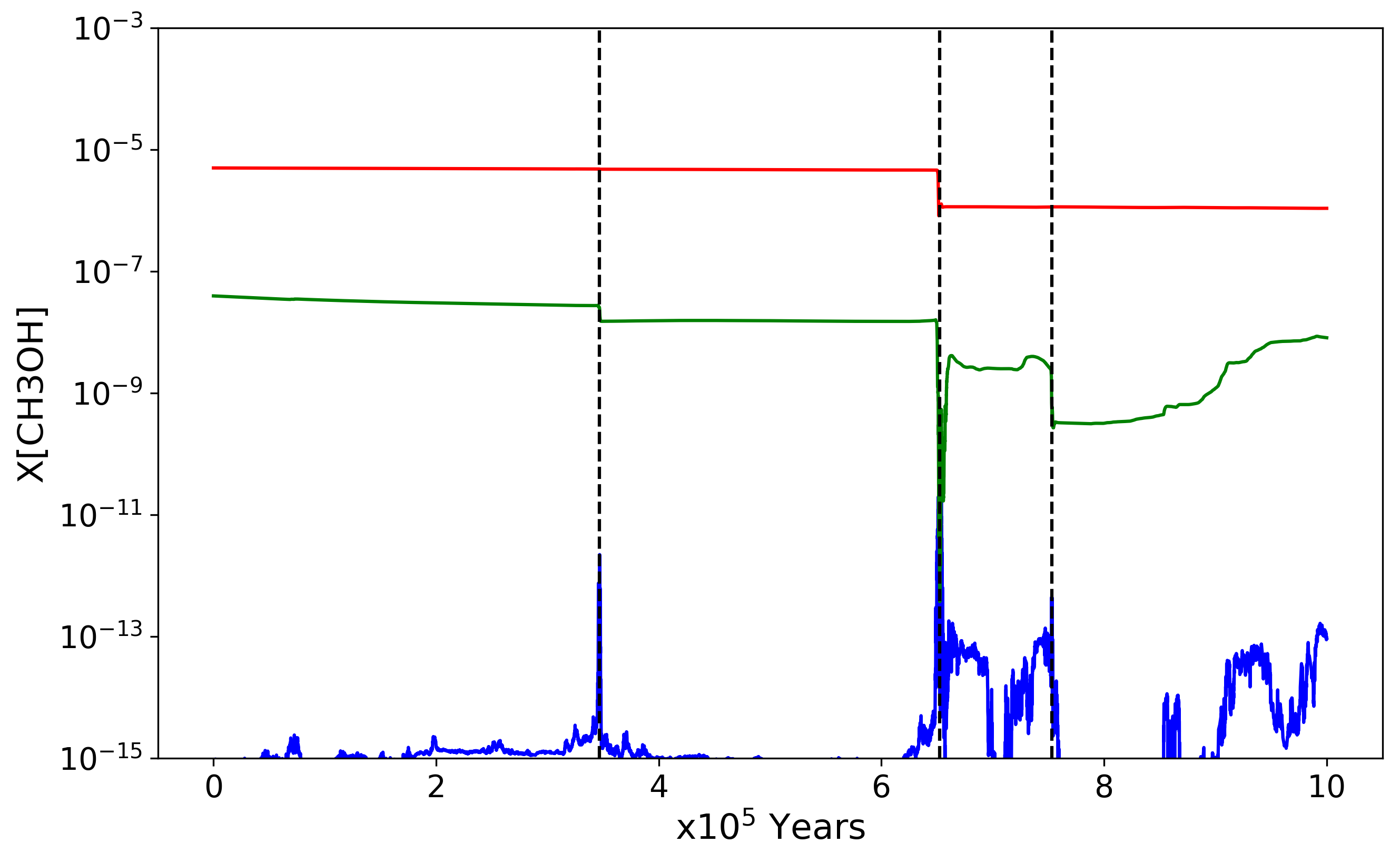}
\end{minipage}
\begin{minipage}[t]{0.3\textwidth}
 \includegraphics[width=5.5cm]{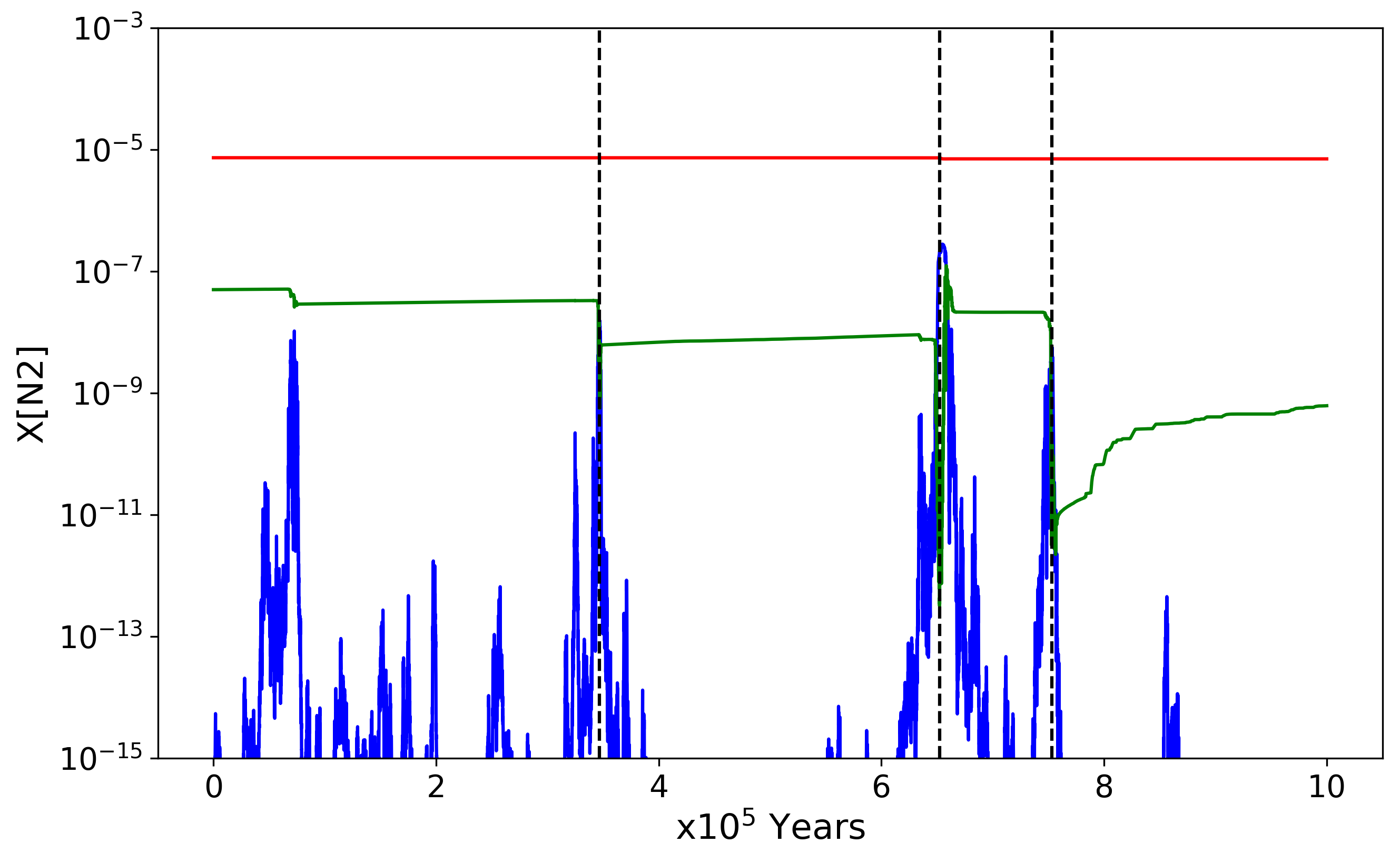}
\end{minipage}
\begin{minipage}[t]{0.3\textwidth}
 \includegraphics[width=5.5cm]{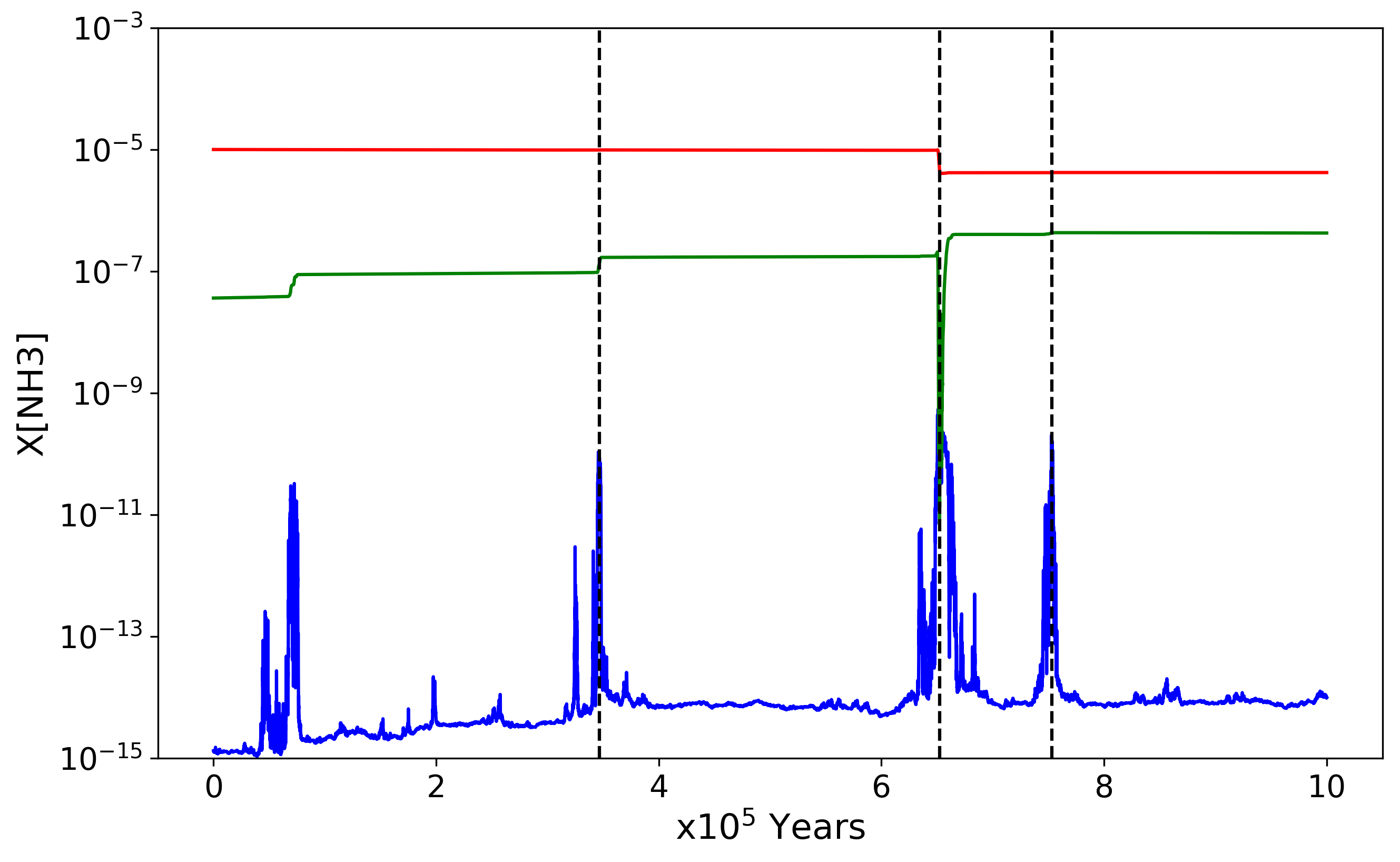}
\end{minipage}
\begin{minipage}[t]{0.3\textwidth}
 \includegraphics[width=5.5cm]{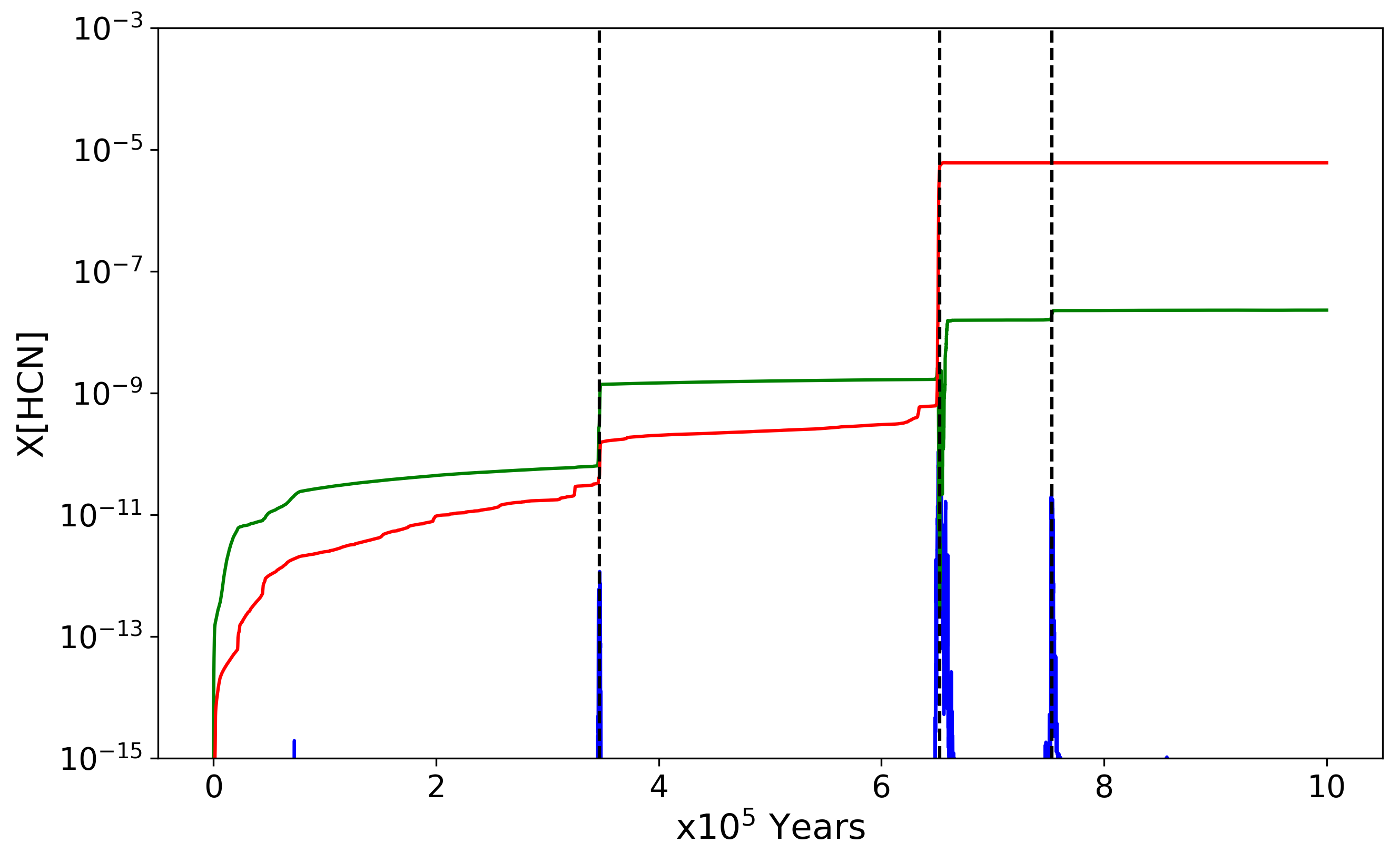}
\end{minipage}
\begin{minipage}[t]{0.3\textwidth}
 \includegraphics[width=5.5cm]{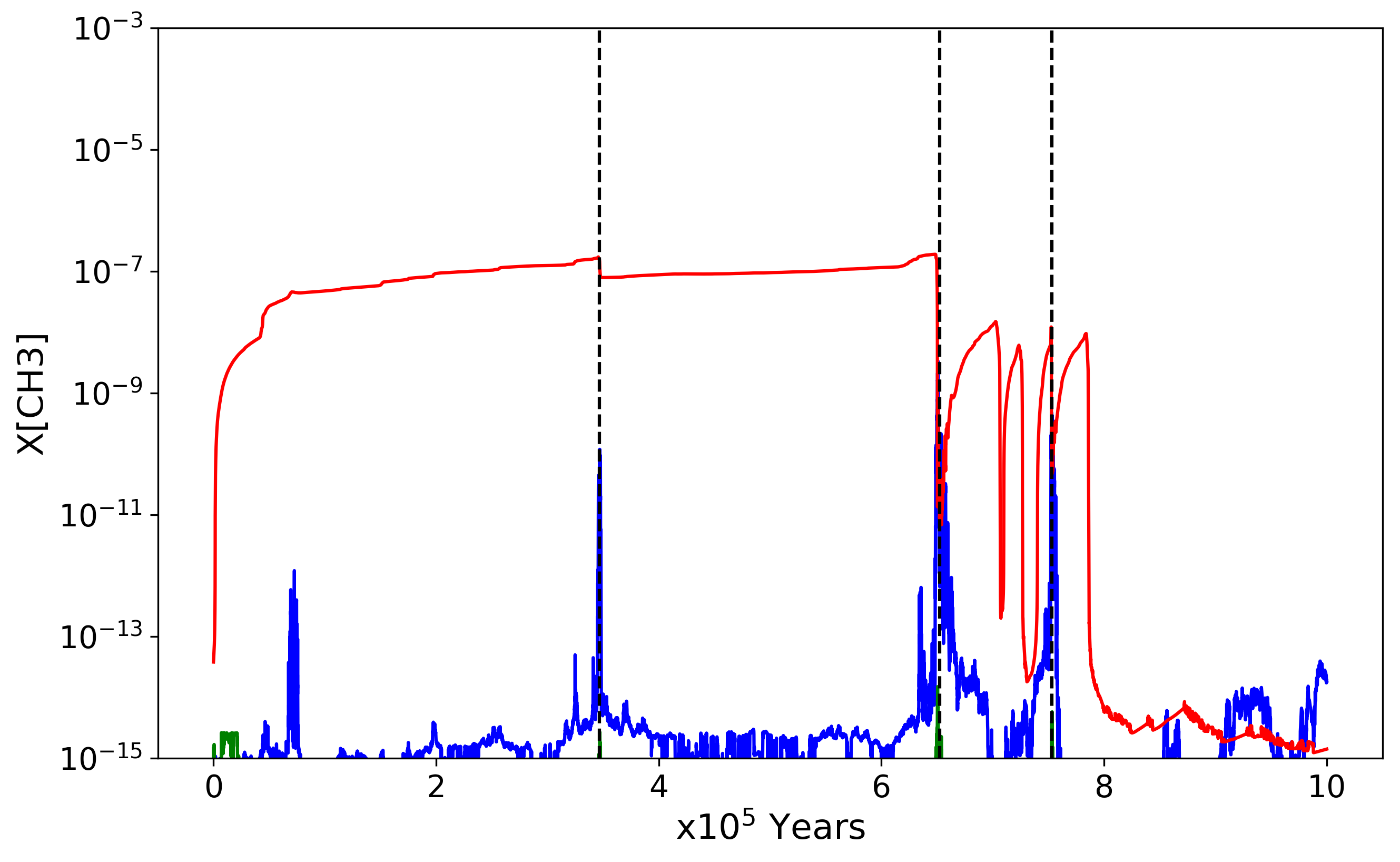}
\end{minipage}
\begin{minipage}[t]{0.3\textwidth}
 \includegraphics[width=5.5cm]{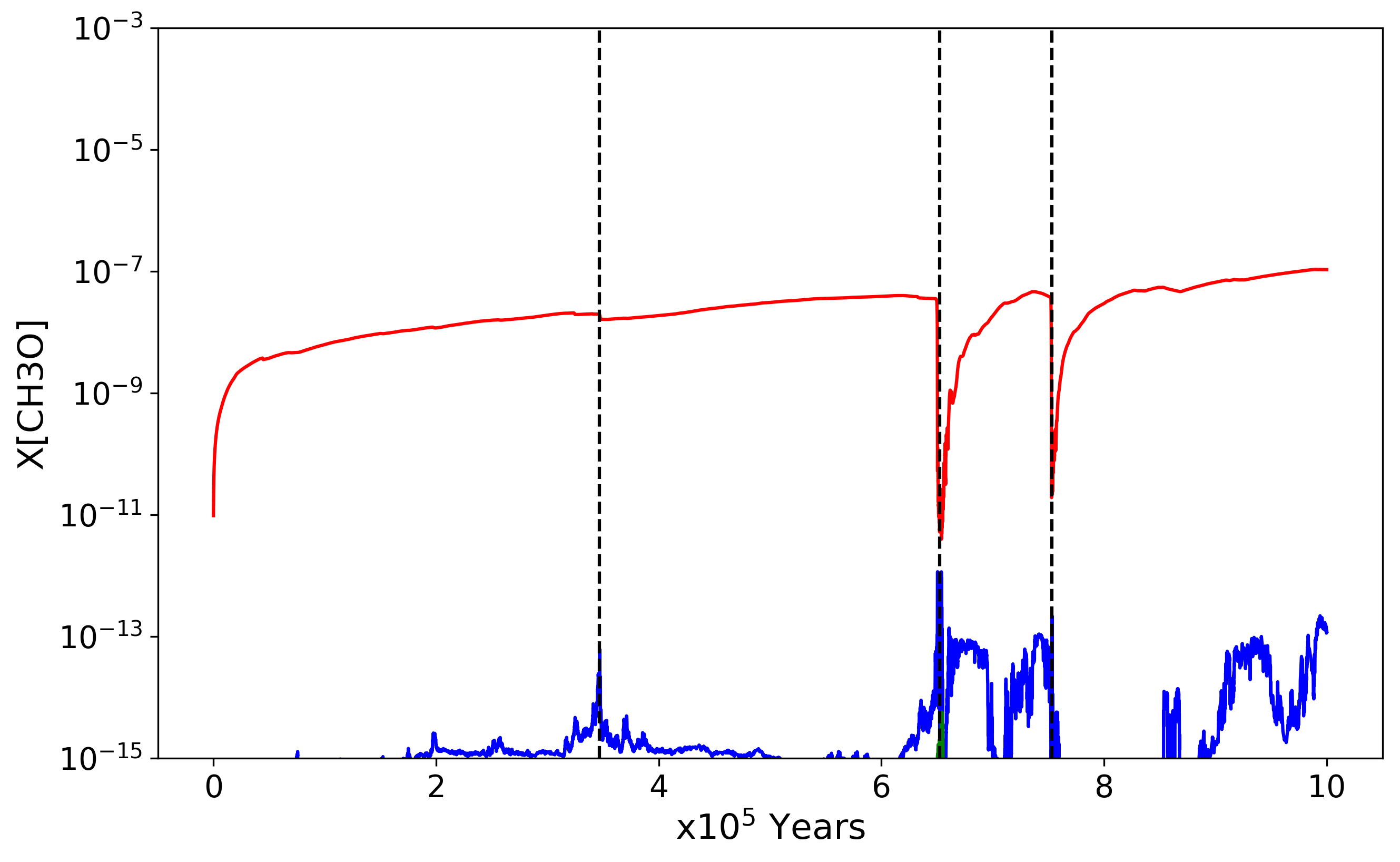}
\end{minipage}
\begin{minipage}[t]{0.3\textwidth}
 \includegraphics[width=5.5cm]{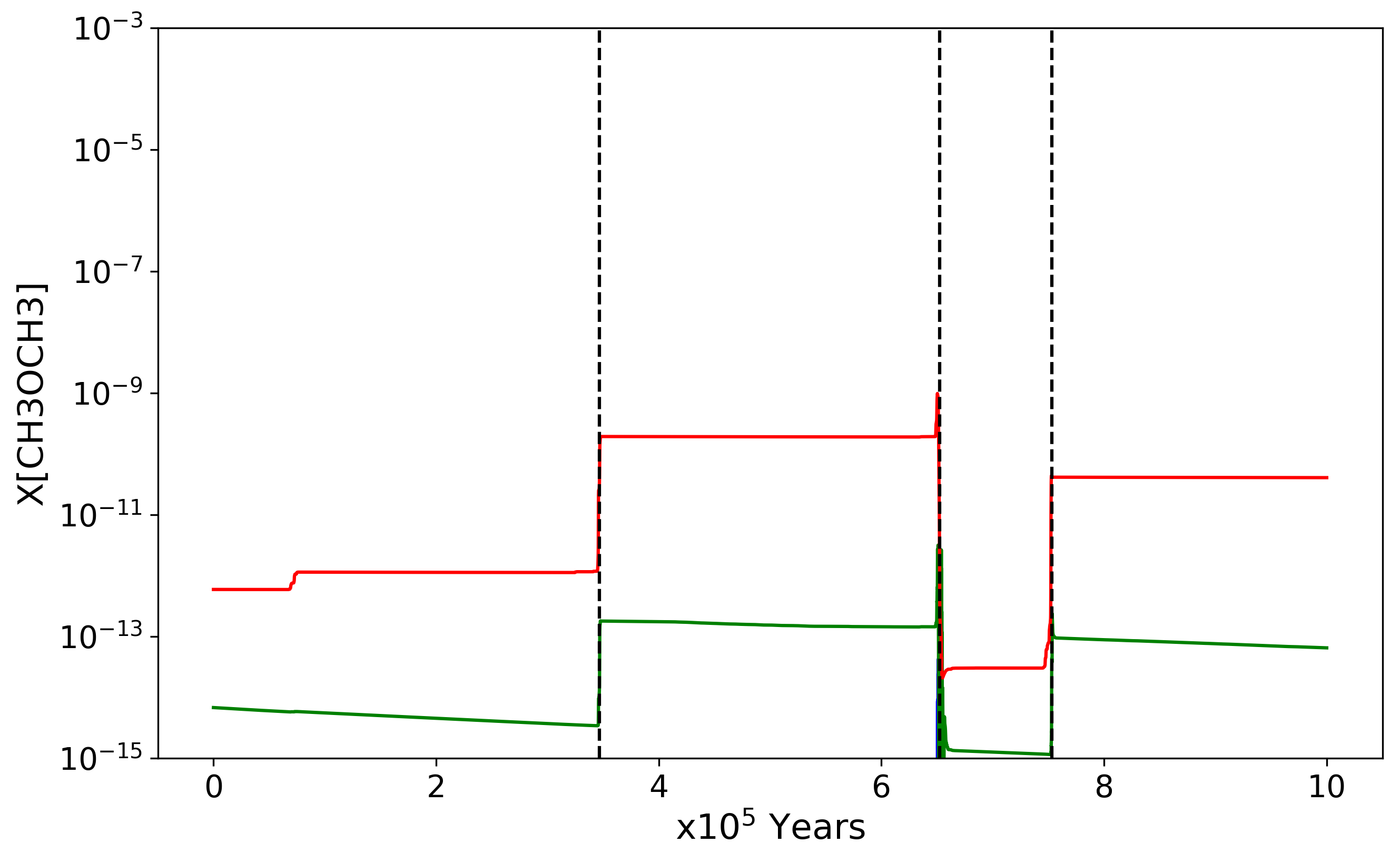}
\end{minipage}
\begin{minipage}[t]{0.3\textwidth}
 \includegraphics[width=5.5cm]{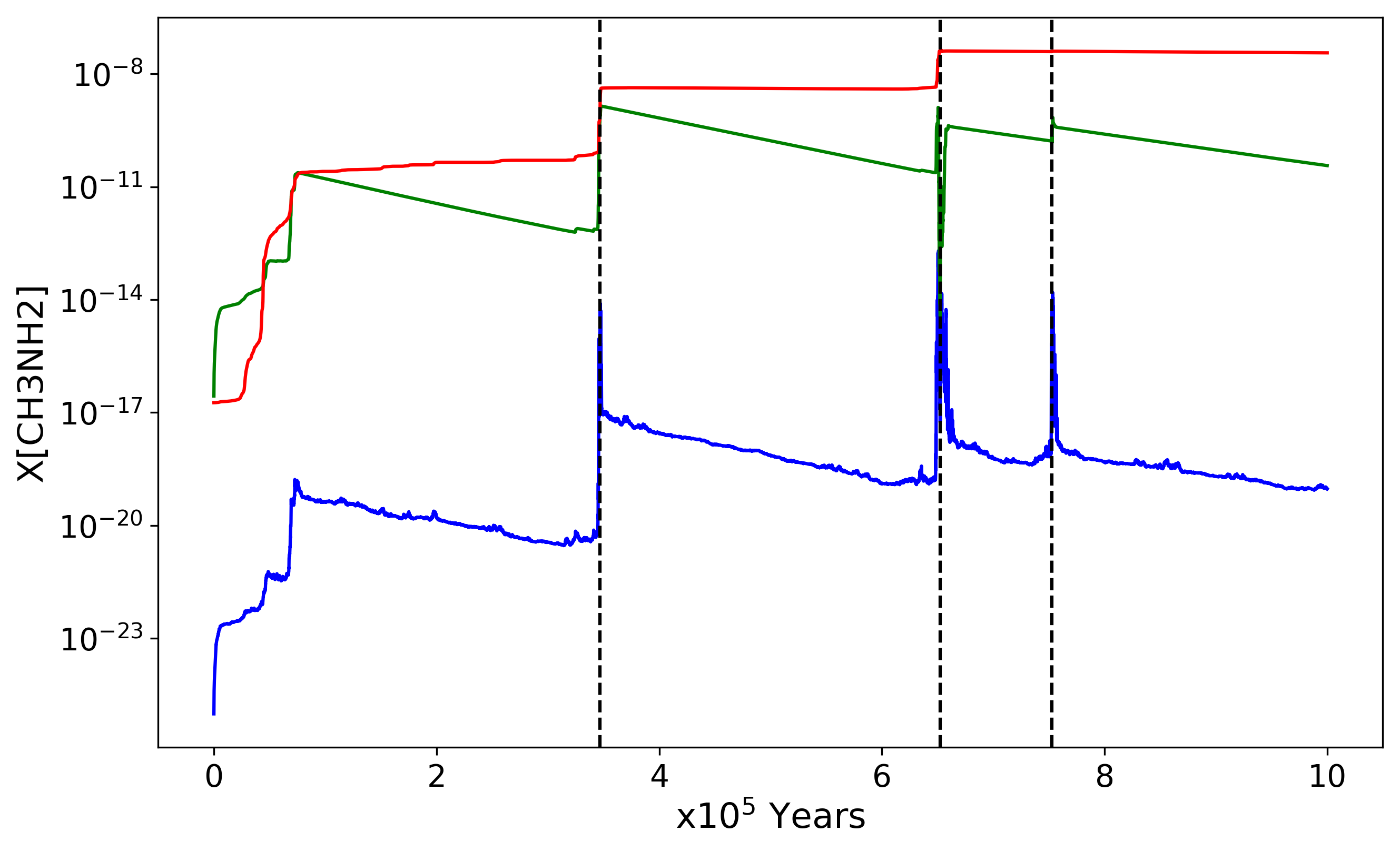}
\end{minipage}
\caption{Temporal variation of dust temperature, UV flux, and the abundances of major species for a selected dust particle. The molecular abundances in the bulk ice mantle (red), on the icy surface (green), and in the gas phase (blue) are shown. Since the dust temperature is mostly $\lesssim 20$ K, the species are mostly in the solid phase, and the gas-phase abundances are much lower than the solid-phase abundances. 
The vertical dashed lines highlight the time steps when the UV flux is at its peak.
}
\label{fig:sample_simulation}
\end{figure*}

The top panels in Figure~\ref{fig:sample_simulation} show the temporal variation of physical parameters.
For most periods, the particle is located near the midplane, where the stellar and interstellar UV radiation are attenuated, and the CR-induced UV photons dominate.
There are, however, three peaks when the stellar UV dominates over CR-induced UV for a short period.
Vertical dotted lines highlight these time steps when the particle is transported near the disk's surface due to turbulent motion, exposing it to high UV flux ($>$10$^{-3}$ G$_{\rm 0}$) and warm ($>$25 K) temperature conditions.
%
%
It's worth noting that the physical characteristics of these three peaks differ significantly. The time scales range from a year to 10$^4$ years. The UV fluence during each event varies over a broad range, from \(10^{-2}\) to \(10^5\)~G$_{\rm 0}$~Years.
%
%

Figure \ref{fig:sample_simulation} also shows the abundances of selected molecules in the sample particle.
The formation of COMs is influenced by the abundances of major molecules containing O, C, and N, e.g., H\(_2\)O, CO, CO\(_2\), CH\(_4\), CH\(_3\)OH, N\(_2\), and NH\(_3\). 
Their abundances undergo sharp changes during periods of intense UV exposure. Notably, over \(10^6\) years, the ice-mantle abundances (red lines in Figure~\ref{fig:sample_simulation}) of CO, CH\(_3\)OH, and NH\(_3\) decrease, whereas that of CO\(_2\) increases slightly. The abundances of H\(_2\)O, CH\(_4\), N\(_2\) in the mantle remain almost constant, sustained by subsequent ice-mantle reactions in post-photodissociation.
Grain surface abundances (green lines in Figure \ref{fig:sample_simulation}) demonstrate pronounced temporal fluctuations compared to the mantle abundances, because the surfaces can interact with the gas. The gas phase abundances, presented by blue lines, show spikes around the time steps of high UV and temperature, due to temporal increase of thermal desorption and photodesorption, as well as enhanced gas-phase reactions of radicals. 

In addition to these species, HCN, which is not initially present in our model, shows a substantial increase in its abundance during the simulation.
Formation of HCN predominantly occurs via gas-phase reactions. CO transforms to H\(_2\)CO through a sequence of hydrogen addition reactions, and a fraction of the produced H$_2$CO (assumed to be 1 \% of the products) is immediately desorbed to the gas phase due to the chemical desorption. N\(_2\) sublimates even at low temperatures; once the temperature exceeds 25 K, its gaseous abundance increases to $\ge$10$^{-8}$. With UV exposure, H\(_2\)CO and N\(_2\) in the gas phase are photodissociated into HCO and N respectively. They subsequently react to produce HCN: \( \text{N + HCO} \rightarrow \text{HCN + O} \). The newly synthesized HCN then re-adsorbs onto the solid phase.

Regarding the formation of COMs, CH\(_3\)OCH\(_3\) and CH\(_3\)NH\(_2\) are chosen as representative O- and N-containing COMs, respectively.
CH\(_3\)OCH\(_3\) is produced via the grain-surface and ice-mantle reactions of CH\(_3\) and CH\(_3\)O \citep{Garrod08}. Given that radicals are formed under UV radiation and their subsequent reactions proceed via thermal diffusion, both high UV intensity and grain temperature are crucial for the formation of COMs. The first UV peak sees a rapid production of CH\(_3\)OCH\(_3\), increasing its abundance on both the grain surface and mantle. However, during the second UV peak, the abundance of CH\(_3\)OCH\(_3\) declines, due to photodissociation. CH\(_3\)NH\(_2\), formed through the CH\(_3\) + NH\(_2\) reaction, on the other hand, sees an increase even during the second UV exposure. As we will discuss in the subsequent sections, the disparity between O-bearing and N-bearing COMs can be attributed to differences in their formation pathways and the behavior of their precursors.
Specifically, conversion of carbon to CO$_2$ ice results in the reduction of O-bearing COMs.

While we plotted CH$_3$OCH$_3$ in Figure~\ref{fig:sample_simulation} as a representative O-bearing COM, its abundance could be overestimated in our model. \cite{Enrique-Romero22} pointed out that the reaction between CH$_3$ and CH$_3$O has two channels: the formation of CH$_3$OCH$_3$ and the formation of CH$_4$ and H$_2$CO. While the latter branch has a higher activation energy ($\sim$1140~K) than the former ($\sim$370~K), it competes with the former branch at low temperatures where the quantum tunneling becomes important. Our model assumes no activation barriers for the former channel. Even if we adopt the activation barrier of $\sim$370 K, the former channel is effectively barrier-less due to the reaction-diffusion competition, since the activation energy for surface diffusion of CH$_3$ is set to $\sim$640~K. Our model, however, does not include the latter channel, and thus could overestimate the production rate of CH$_3$OCH$_3$ at the low temperature. Further studies such as the evaluation of the barrier width to calculate the tunneling rate are required to improve the accuracy of the network model of COMs.

\subsection{Effect of Stochastic UV Radiation}
We investigate the dependence of the COM abundances on the UV fluence ($\Gamma$) and the UV flux based on the calculations of the 1000 particles with the size of 1~$\mu$m.
Since COMs are mainly formed by the reactions between radicals in ice, which are formed by photodissociation of the major ice molecules, we present the abundances of major carriers of C, N, and O before describing the COM chemistry.

\subsubsection{The main carriers of C, N, and O} \label{sec:main_carrier}
\begin{figure*}
 \centering
\begin{minipage}[t]{0.3\textwidth}
 \includegraphics[width=5.5cm]{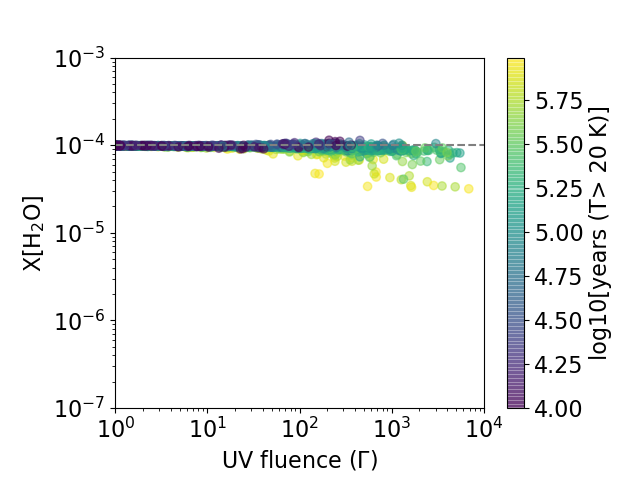}
\end{minipage}
\begin{minipage}[t]{0.3\textwidth}
 \includegraphics[width=5.5cm]{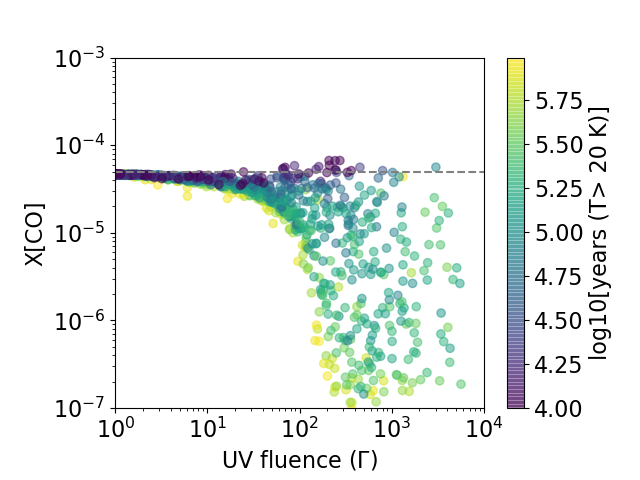}
\end{minipage}
\begin{minipage}[t]{0.3\textwidth}
 \includegraphics[width=5.5cm]{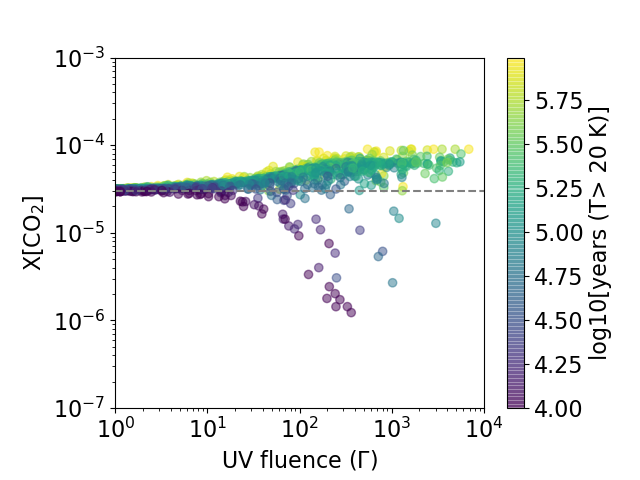}
\end{minipage}
\begin{minipage}[t]{0.3\textwidth}
 \includegraphics[width=5.5cm]{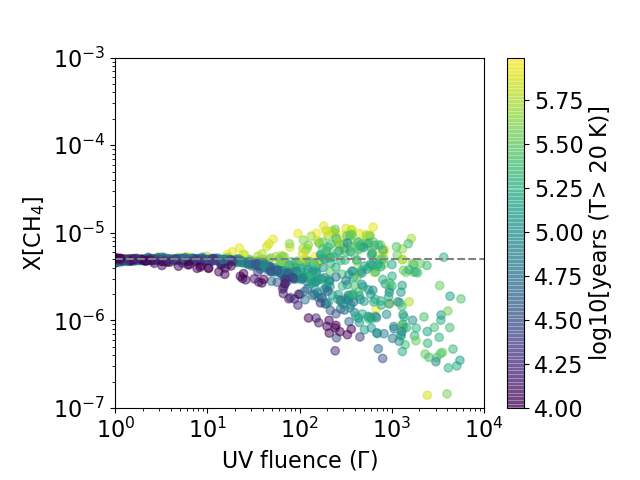}
\end{minipage}
\begin{minipage}[t]{0.3\textwidth}
 \includegraphics[width=5.5cm]{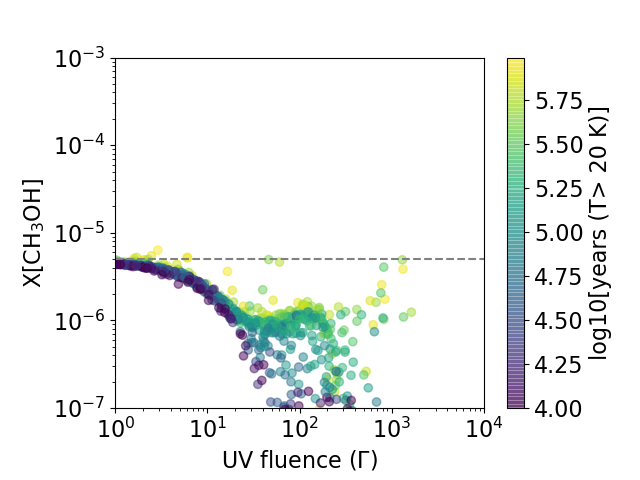}
\end{minipage}
\begin{minipage}[t]{0.3\textwidth}
 \includegraphics[width=5.5cm]{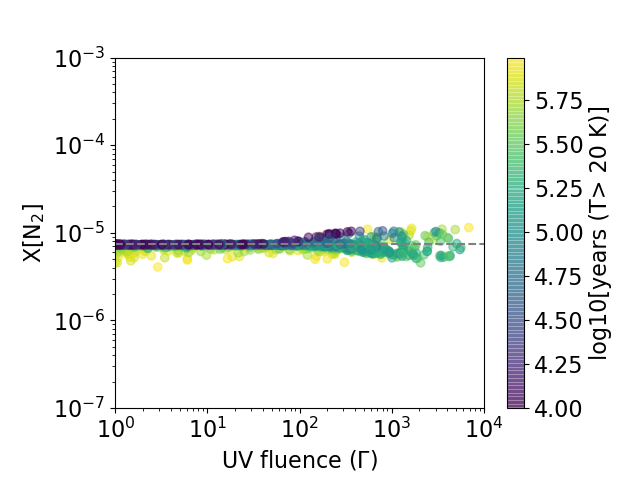}
\end{minipage}
\begin{minipage}[t]{0.3\textwidth}
 \includegraphics[width=5.5cm]{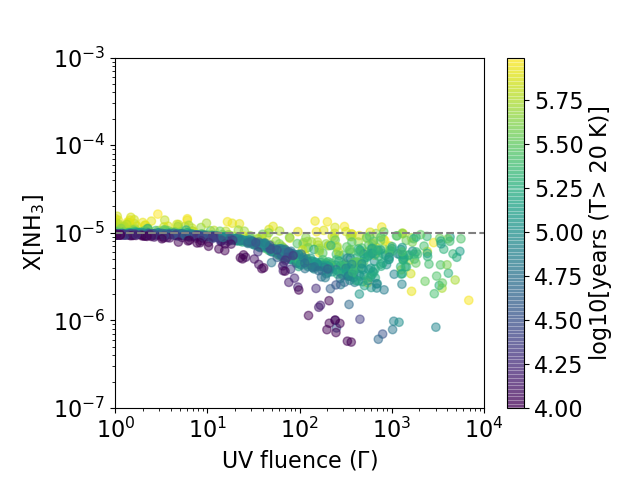}
\end{minipage}
\begin{minipage}[t]{0.3\textwidth}
 \includegraphics[width=5.5cm]{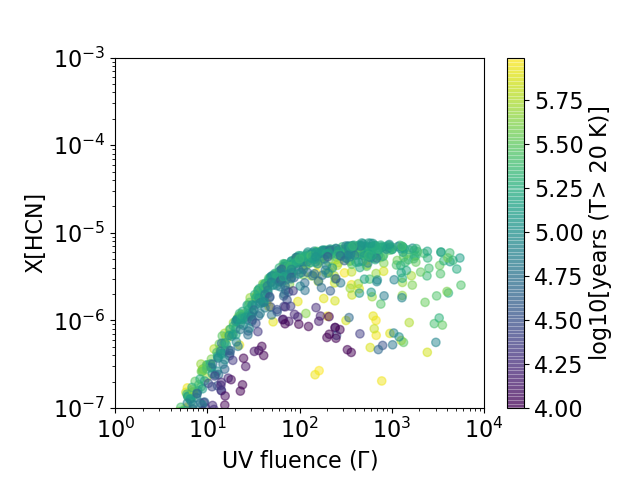}
\end{minipage}
\caption{Final abundances of major molecules containing C, N, and O as functions of $\Gamma$.
The color scale indicates the cumulative duration when the dust temperature is above 20~K, which serves as a measure of the efficiency of the dust surface reaction because the thermal diffusion of radicals becomes efficient above $\sim$20 K.
The dotted horizontal lines indicate the initial abundances.
}
\label{fig:CNO_carriers}
\end{figure*}

Figure~\ref{fig:CNO_carriers} shows the abundance of the major molecules containing C, N, and O at 10$^6$ yr as functions of \(\Gamma\). For particles which fall inside 1~au during the simulation, we plot the abundances at 1~au.
We summed abundances in the gas phase, on the grain surface, and in the bulk ice mantle.
The total duration time when the dust temperature is above 20~K is depicted by a color scale as a measure of the efficiency of the dust surface reaction via thermal diffusion. This threshold temperature is chosen by considering the diffusion rates of small radicals, such as CH\(_3\), which contribute to the formation of COMs.
The diffusion rate of adsorbed species is sensitively dependent on the dust temperature. A radical contributes efficiently to COM formation when its diffusion rate is higher (or equivalently, the diffusion timescale is shorter) than the formation rate of the radical (i.e., photodissociation rate of precursors). For small radicals such as CH\(_3\), the threshold temperature is around $\sim$20~K in our fiducial model.

The main carriers of carbon at the end of the simulations are CO, CO\(_2\), CH\(_4\), and CH\(_3\)OH, all of which are assumed to be abundant in initial conditions.
In the initial condition, CO is the most abundant carbon reservoir.
The decline of the CO abundance starts when \(\Gamma\) exceeds \(10\), but the extent of the decline depends on the thermal history of each particle. The particles that have peak dust temperatures below 20~K show little or no decline in the CO abundance. In contrast, particles that spend approximately $10^5$ years above 20~K show a decline in the CO abundance by around three orders of magnitude.
The photodissociation of H\(_2\)O ice, which is the dominant oxygen reservoir in the ice mantle, produces OH radicals.
The reaction of OH radicals with CO is very efficient at $\gtrsim$ 20~K due to the enhanced diffusion of CO, and most of CO is converted and locked to CO\(_2\).
While CO$_2$ is also photodissociated, the photodissociation product, CO, is converted back to CO$_2$ by the reaction with OH. 

The abundance of CH\(_3\)OH decreases when \(\Gamma\) exceeds \(10\), showing a decline by more than an order of magnitude.
When CH\(_3\)OH is photodissociated, $\sim$70 \% forms CH\(_2\)OH (CH\(_3\)OH $\rightarrow$ CH\(_2\)OH + H), while the remaining 30 \% splits equally between CH\(_3\)O and CH\(_3\) in our chemical network. CH\(_2\)OH can either revert back to CH\(_3\)OH via the hydrogenation reaction or further dissociate to H\(_2\)CO (CH\(_2\)OH + H $\rightarrow$ H\(_2\)CO + H\(_2\)). Subsequent photodissociation of H\(_2\)CO yields HCO and CO. Similarly, HCO is then converted back to CH\(_2\)OH via hydrogenation or photodissociated to CO. This intricate interplay of photodissociation and hydrogen addition reactions results in carbon going back and forth between CH\(_3\)OH and CO in the chemical network under the influence of UV radiation.
The abundance of CH\(_3\)OH is not only influenced by the amount of UV radiation but also depends significantly on the dust temperature. Particles that experience prolonged periods with dust temperatures above 20~K show a higher final abundance of CH\(_3\)OH. This is attributed to the efficient diffusion of CH\(_3\) radicals at around 20~K, which enhances the formation of CH\(_3\)OH through the CH\(_3\) + OH reaction on grains, in addition to the hydrogenation of CO at lower temperatures.

CH$_4$, on the other hand, exhibits an increase in its abundance with increasing $\Gamma$.
This is attributed to the photodissociation of CH$_3$OH and other molecules containing CH$_3$ groups, generating CH$_3$, which then undergo subsequent hydrogenation processes to form CH$_4$. As mentioned earlier, while CH$_3$ is not a major product of CH$_3$OH photodissociation in our chemical network, it does contribute to the formation of CH$_4$ via hydrogenation.

\begin{figure*}
 \includegraphics[width=14cm]{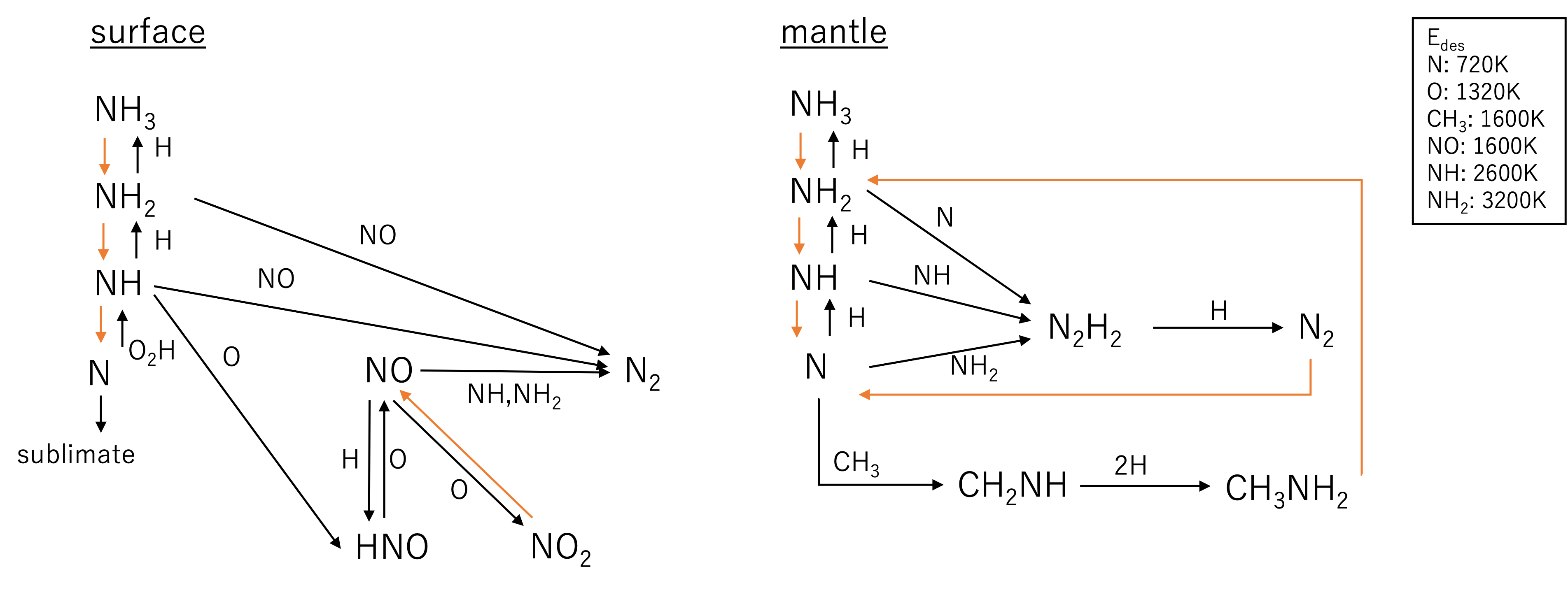}
\caption{The chemical reactions related to the NH$_3$ and N$_2$ chemistry on grain surfaces. The orange arrows represent the photodissociation processes. The binding energy of key species is shown on the top right.
}
\label{fig:N-chart}
\end{figure*}

The overall abundances of N$_2$ and NH$_3$ do not change significantly over \(10^6\) years, and N$_2$ and NH$_3$ ices are dominant nitrogen carriers even under strong UV irradiation ($\Gamma > 100$). This result contrasts with carbon and oxygen chemistry, where carbon and oxygen are eventually locked in CO$_2$ ice (and H$_2$O ice).
We depict reactions relevant to N-bearing molecules in Figure~\ref{fig:N-chart}.
Under UV radiation, NH$_3$ ice experiences a sequence of photodissociation, producing radicals such as NH$_2$, NH, and atomic N. These radicals can be converted to NH$_3$ again through a sequence of hydrogenation addition reactions. The photodissociation of CO$_2$ generates CO and atomic O, the latter of which reacts with atomic N to form NO; a subsequent reaction with atomic O leads to NO$_2$. However, photodissociation of NO$_2$ forms NO again, allowing NO and NO$_2$ to coexist. Given the low desorption energy of NO (1600 K) and thus the low activation energy for surface diffusion, it can diffuse on grain surfaces even at around 20 K and react with NH or NH$_2$ to form N$_2$:
\begin{align}
\mathrm{NH} + \mathrm{NO} & \rightarrow \mathrm{N}_2 + \mathrm{H} + \mathrm{H}, \\
\mathrm{NH}_2 + \mathrm{NO} & \rightarrow \mathrm{N}_2 + \mathrm{H}_2\mathrm{O}.
\end{align}
N$_2$ is also photodissociated, producing atomic N, which is converted to NH$_3$ or N$_2$ again. 

The reactions involving NH$_3$ and N$_2$ also occur within the ice mantle, but the chemical pathways differ from those on grain surfaces, primarily due to the slower diffusion rates in the ice mantle than on the surface.
Atomic H and N have desorption energy of 440 and 720~K, respectively, and they still diffuse and react within the mantle. However, species such as atomic O, NO, NH, and NH$_2$, which play important roles in grain surface chemistry, do not diffuse effectively in the mantle. 
As a result, after the photodissociation of NH$_3$, the reaction between N and NH$_2$ leading to the formation of N$_2$H$_2$ becomes more important. N$_2$H$_2$ reacts with atomic H, forming N$_2$:

\begin{align}
\mathrm{N} + \mathrm{NH}_2 & \rightarrow \mathrm{N}_2\mathrm{H}_2, \\
\mathrm{N}_2\mathrm{H}_2 + \mathrm{H} & \rightarrow \mathrm{N}_2 + \mathrm{H}_2 + \mathrm{H}.
\end{align}
Furthermore, the photodissociation of N$_2$ and NH$_3$ in the mantle produces atomic N, which can react with CH$_3$ to form CH$_2$NH. Then CH$_2$NH undergoes a hydrogenation process to form CH$_3$NH$_2$, a primary N-bearing COMs in our fiducial model. It is noted that we assume no activation energy barrier for hydrogenation addition reactions to CH$_2$NH.

Exceptionally, the HCN abundance increases up to $\sim 10^{-6}$ when the UV fluence exceeds $\Gamma \sim 10$.
The formation pathway of HCN starts with the thermal or non-thermal desorption of N\(_2\) ice, followed by photodissociation producing atomic N. Then atomic N reacts with hydrocarbons in the gas phase. For instance, in addition to the reaction described in section~\ref{sec:abundances_example}, a reaction N + CH\(_3\) $\rightarrow$ H\(_2\)CN + H, followed by a reaction H\(_2\)CN + H $\rightarrow$ HCN + H\(_2\), contributes significantly to the production of HCN.
The abundance of HCN on grain surface is primarily determined by the adsorption of gas phase HCN, and the contribution of grain surface reactions are negligible.

\subsubsection{Complex Organic Molecules} \label{sec:coms}
In this subsection, we discuss the abundance of COMs.
We found that UV radiation has both positive and negative effects on COMs abundances. The photodissociation of abundant stable molecules produces radicals, which thermally diffuse and react with each other to form COMs.
On the other hand, the UV radiation photodissociates and destroys COMs.
Therefore the temporal variation of the UV flux and dust temperature due to particle motion in the turbulent disk is important for the formation and destruction of COMs.

\begin{figure*}
 \centering
\begin{minipage}[t]{0.3\textwidth}
 \includegraphics[width=5.5cm]{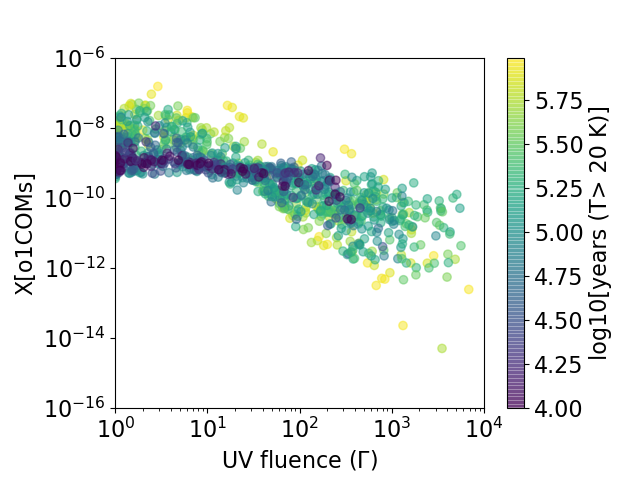}
\end{minipage}
\begin{minipage}[t]{0.3\textwidth}
 \includegraphics[width=5.5cm]{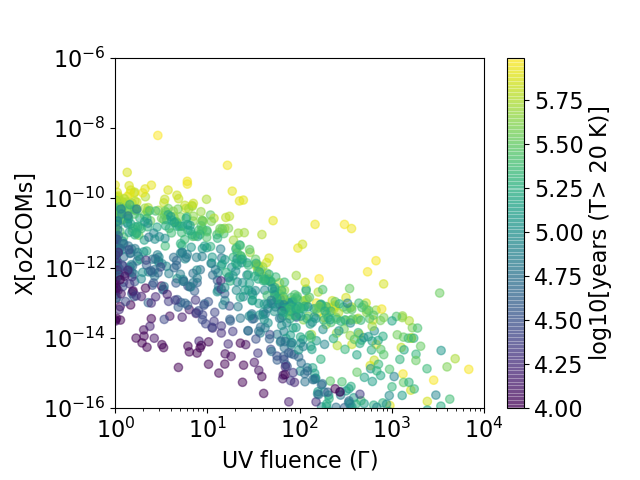}
\end{minipage} \\
\begin{minipage}[t]{0.3\textwidth}
 \includegraphics[width=5.5cm]{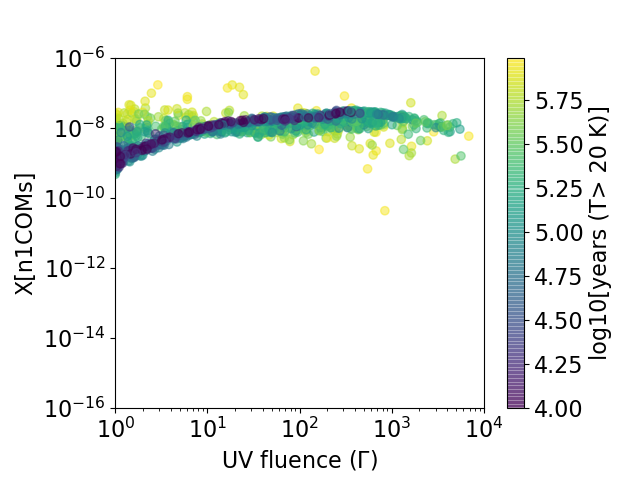}
\end{minipage}
\begin{minipage}[t]{0.3\textwidth}
 \includegraphics[width=5.5cm]{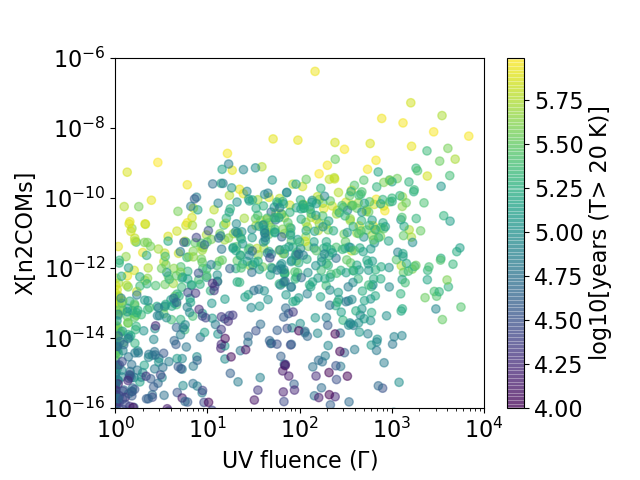}
\end{minipage}
\caption{Same as Figure~\ref{fig:CNO_carriers} but for COMs of four groups as functions of $\Gamma$: COMs containing one or more O (o1COMs), two or more O (o2COMs), one or more N (n1COMs), and two or more N (n2COMs). The colors indicate the cumulative duration when dust temperature is $\ge 20$ K. 
}
\label{fig:COMs}
\end{figure*}

Our chemical network contains 365 species of COMs in the gas phase, including both neutral ones and ions, and an additional 147 COMs in the solid phase (i.e. grain srface and ice mantle). As discussed below, O-bearing COMs and N-bearing COMs exhibit different dependencies on the UV fluence. We therefore present the total abundances of COMs containing one or more oxygen atoms (o1COMs), two or more oxygen atoms (o2COMs), one or more nitrogen atoms (n1COMs), and two or more nitrogen atoms (n2COMs) as a function of the normalized UV fluence, $\Gamma$ (Figure~\ref{fig:COMs}).
Figure~\ref{fig:COMs} shows the sum of the abundances in the gas and solid phases as in Figure~\ref{fig:CNO_carriers}.
Most COMs are frozen in the ice mantle for the range of temperatures covered by our calculations (mostly $<$100 K).
Although various COMs have been detected in the observations toward the pre-disk stage, i.e. prestellar and protostellar cores, our initial conditions include only one COM, CH$_3$OH, so that we can evaluate which COMs are formed within the disk in our fiducial model (see also Section \ref{sec:inherit}).
Since CH$_3$OH is by far the most abundant O-bearing COM, we exclude CH$_3$OH from o1COMs in order to focus our attention on the newly formed COMs inside the disk.

Various O-bearing COMs are newly formed inside the disk.
For each particle, we listed the species whose abundance is higher than the 1~$\%$ of the sum of o1COMs. Notably, CH$_3$OCH$_3$, CH$_3$CHO, and C$_2$H$_5$OH appear in most particles.
The formation of CH$_3$OCH$_3$ is predominantly a result of the reaction between CH$_3$ and CH$_3$O radicals. Similarly, C$_2$H$_5$OH is synthesized through the reaction between CH$_3$ and CH$_2$OH radicals. As for CH$_3$CHO, its formation involves a two-step process: initially, CH$_2$ reacts with CHO to form CH$_2$CHO, followed by the reaction CH$_2$CHO + H $\rightarrow$ CH$_3$CHO.
CH$_3$COCH$_3$ and CH$_3$CONH$_2$ exceed the 1~$\%$ abundance threshold in more than 500 particles, while CH$_3$CO, CH$_2$OHNH$_2$, C$_2$H$_5$CHO, and CH$_3$ONH$_2$ appearing in some particles.
In the similar analysis for o2COMs, HCOOCH$_3$, OHCCHO, and CH$_2$OHCHO are abundant in most particles, and 
CH$_3$CO(CH$_2$OH) appears in about half of the particles.
A characteristic feature of O-bearing COMs is that their abundances decrease as the UV fluence increases.
This means that the negative effect of UV radiation is greater than the positive effect for the abundance of O-bearing COMs.
A sequence of the photodissociation of O-bearing COMs often produces CO in our model.
As discussed in the previous subsection, CO reacts efficiently with OH and is converted to CO$_2$ in the ice mantle.
Therefore, the total abundance of O-bearing COMs decreases with the UV fluence. For particles with a similar UV fluence, O-bearing COMs are more abundant in particles with a longer cumulative duration of $T \gtrsim 20$ K, because the thermal diffusion of CO to form COMs requires the warm temperature.

%

Among the n1COMs, CH$_3$NH$_2$ is the most abundant in our simulations, and the total abundance of n1COMs is almost equivalent to that of CH$_3$NH$_2$.
For other n1COMs, C$_2$H$_5$CN, C$_2$H$_3$CN, and CH$_3$CHCN are formed with an abundance exceeding 1 \% of the total n1COMs abundance in a few tens of particles. These molecules exhibit behavior similar to HCN; their abundances increase when $\Gamma$ exceeds 3. This trend is attributed to adsorbed HCN undergoing photodissociation, subsequently forming these COMs from the photodissociation product, CN radicals.

The effect of UV is more obvious for n2COMs. Among n2COMs, NH$_2$-NH$_2$ has the highest abundance, and the total n2COMs abundance is almost equivalent to that of NH$_2$-NH$_2$. The abundance of the other n2COMs is less than 1 \% of the total n2COMs in most particles. However, an exception is observed in 51 particles where NH$_2$CNH$^+$ shows an abundance exceeding 1 \%. This is attributed to the reaction pathways that are efficient in regions of strong UV radiation and higher temperatures. The reactions proceed as follows in the gas phase:
\begin{align}
\mathrm{NH}_3 + \mathrm{CN} & \rightarrow \mathrm{NH}_2\mathrm{CN} + \mathrm{H}, \\
\mathrm{NH}_2\mathrm{CN} + \mathrm{H}_3^+ & \rightarrow \mathrm{NH}_2\mathrm{CNH}^+ + \mathrm{H}_2.
\end{align}


The stark contrast between O-bearing COMs and N-bearing COMs is due to the formation and destruction of major O-bearing and N-bearing species described in the previous subsection. While CO is converted and locked to CO$_2$ with high UV fluence, N$_2$ and NH$_3$ remain abundant in most of the particles and contribute to the formation of N-bearing COMs.

\subsubsection{Model with time-averaged UV flux} \label{sec]time-average}
While O-bearing COMs and N-bearing COMs show the general trend of decreasing and increasing with the UV fluence, respectively, their abundances vary significantly among particles even with a similar UV fluence. The variation is at least partly due to the different thermal histories of each particle; i.e., the COM abundances tend to be higher in particles with a longer cumulative time of $T_{\rm dust} \gtrsim 20$ K. 
In response to conventional assumption that COM production is primarily determined by the total amount of UV, we clarify the importance of time-variable UV flux, by performing a time-averaged UV flux model; for each particle, we derive the time-averaged UV flux by dividing the UV fluence by 10$^6$ year (or the time before it falls inside 1~au from the central star).
The time variation of other physical parameters such as the temperature and density are assumed to be the same as in the fiducial model.
The COM abundances should be similar between the fiducial model and the time-averaged UV flux model, if the conventional assumption is valid.

\begin{figure*}
 \centering
\begin{minipage}[t]{0.3\textwidth}
 \includegraphics[width=5.5cm]{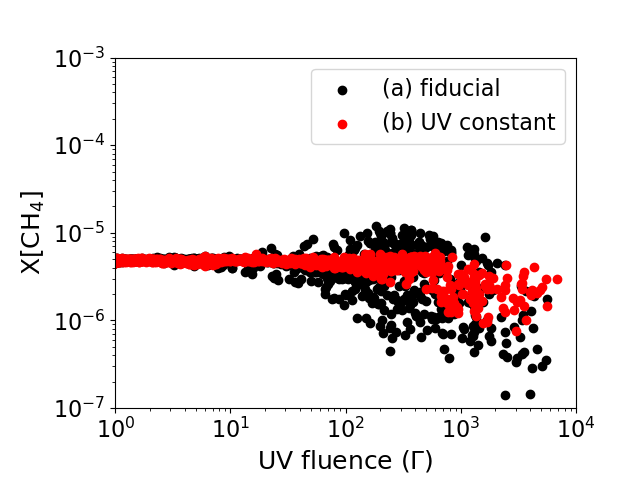}
\end{minipage}
\begin{minipage}[t]{0.3\textwidth}
 \includegraphics[width=5.5cm]{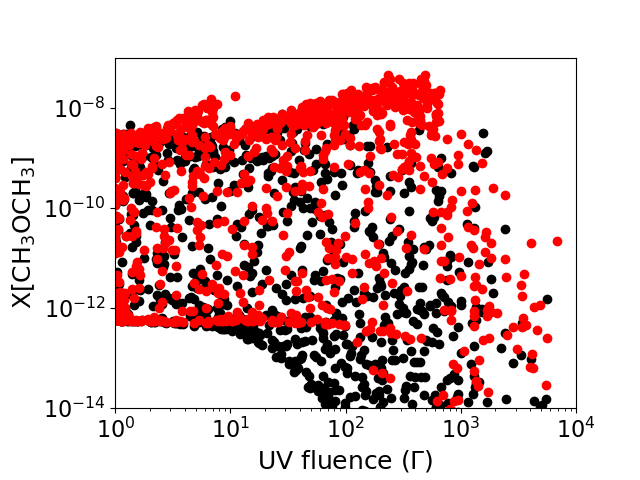}
\end{minipage}
\begin{minipage}[t]{0.3\textwidth}
 \includegraphics[width=5.5cm]{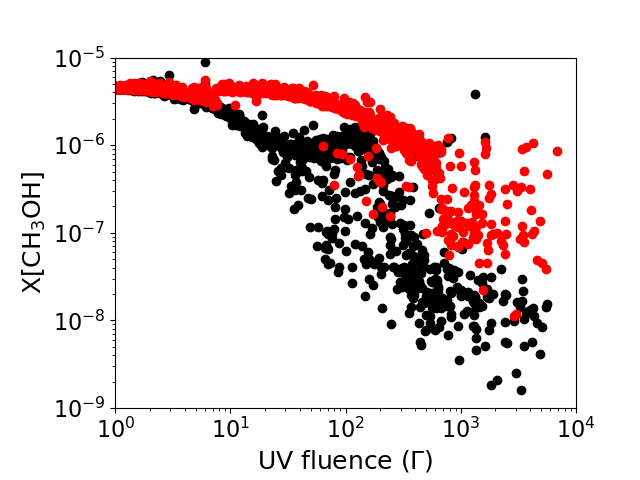}
\end{minipage}
\begin{minipage}[t]{0.3\textwidth}
 \includegraphics[width=5.5cm]{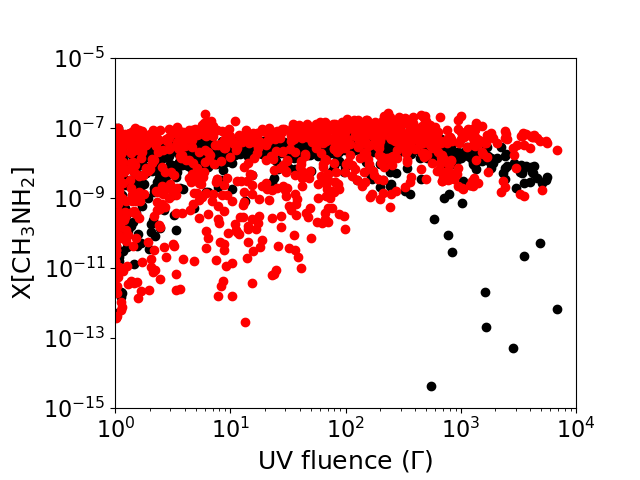}
\end{minipage}
\caption{Final molecular abundances of selected molecules as functions of $\Gamma$ in the time-averaged UV model (red dots). For comparisons, the corresponding results in the fiducial model are shown by black dots. 
}
\label{fig:compare_UV}
\end{figure*}

Figure~\ref{fig:compare_UV} compares the abundances of \( \text{CH}_4 \), \( \text{CH}_3\text{OH} \), \( \text{CH}_3\text{OCH}_3 \), and \( \text{CH}_3\text{NH}_2 \) between the fiducial model and the time-averaged UV model.
The abundance of \( \text{CH}_3\text{OH} \) and \( \text{CH}_3\text{OCH}_3 \) are higher in the time-averaged UV model than in the fiducial model. A naive interpretation is that the stochastic strong UV radiation dissociates the COMs, while such events do not exist in the time-averaged UV model (see \S \ref{sec:threshold}).
This effect can also be applied to CH$_4$; the number of particles with low CH$_4$ abundance (e.g. $\le 10^{-6})$ is reduced in the time-averaged UV model. For some particles, however, CH$_4$ abundance is higher in the fiducial model, because strong UV can photodissociate CO and CO$_2$ to produce C atom, which is then converted to CH$_4$.
The abundance of \( \text{CH}_3\text{NH}_2 \), on the other hand, displays greater variability in the time-averaged UV and tends to be lower at least in some particles.
The variation in the CH$_3$NH$_2$ abundance in the time-averaged UV model can be attributed to the specific conditions required for the formation of its precursor, HCN. As discussed in Section \ref{sec:main_carrier}, HCN is formed in the gas phase through a series of chemical reactions. The formation pathways become more efficient when dust temperatures are high enough for the relevant molecules to be thermally desorbed into the gas phase and be exposed to the UV radiation. When UV radiation is time-averaged, the UV fluence during the periods of higher dust temperatures is reduced. This results in a lower rate of HCN formation, which tends to a reduction in the abundance of CH$_3$NH$_2$ in the time-averaged UV model.
Our results suggest that it is important to follow the trajectories of individual particles and consider their chemical evolution accordingly for simulating the COM formation in disks.


\section{Discussion}
\subsection{Threshold UV Fluence for the destruction of COMs}\label{sec:threshold}
The comparison of O-bearing COM abundances between the fiducial model and the time-averaged UV model in Sec. \ref{sec]time-average} indicates that a temporal strong UV irradiation is destructive for COMs compared with continuous, moderate UV irradiation. Then it is useful to clarify the conditions for destructive events for COMs.
To analyze the abundance change induced by the temporal strong UV irradiation of species $i$, we pick up ``UV event'' from the histories of all the particles. The start of each event is defined as a moment when the UV flux exceeds $1\times 10^{-4}$ G$_{\rm 0}$, i.e., more than about an order of magnitude larger than the flux of CR-induced UV photons, and the abundance of species $i$ at this moment is denoted as $X_{\rm 0}(i)$.
We define that the event ends when the UV flux declines to $1\times 10^{-4}$ $G_{\rm 0}$, and the abundance then is denoted as $X_{\rm 1}(i)$. The duration of the event is denoted as $\Delta t$. For each molecule, we select the relevant UV events by setting two conditions (I) $X_{\rm 0}(i)\ge 1\times 10^{-14}$ and (II) $X_{\rm 1}(i)/X_{\rm 0}(i)\ge 2$ or $X_{\rm 1}(i)/X_{\rm 0}(i)\le 0.5$.





\begin{figure*}
 \centering
\begin{minipage}[t]{0.3\textwidth}
 \includegraphics[width=5.5cm]{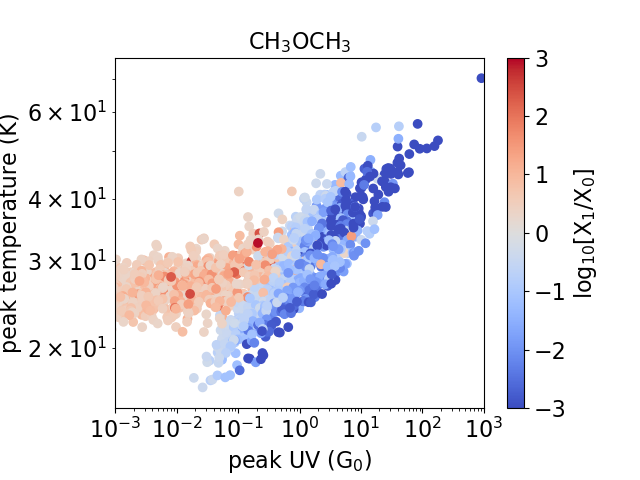}
\end{minipage}
\begin{minipage}[t]{0.3\textwidth}
 \includegraphics[width=5.5cm]{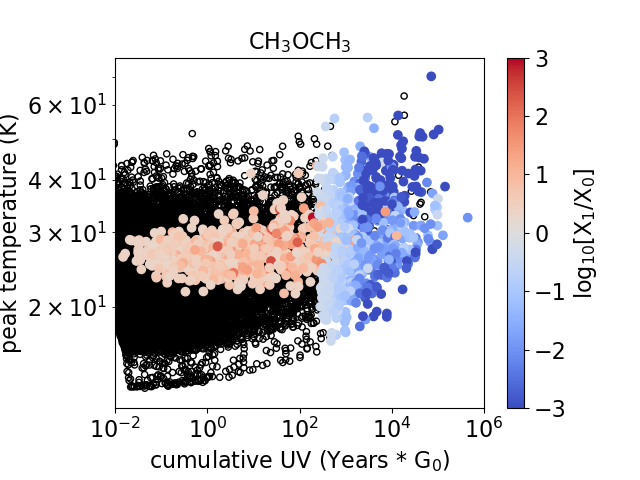}
\end{minipage} \\
\begin{minipage}[t]{0.3\textwidth}
 \includegraphics[width=5.5cm]{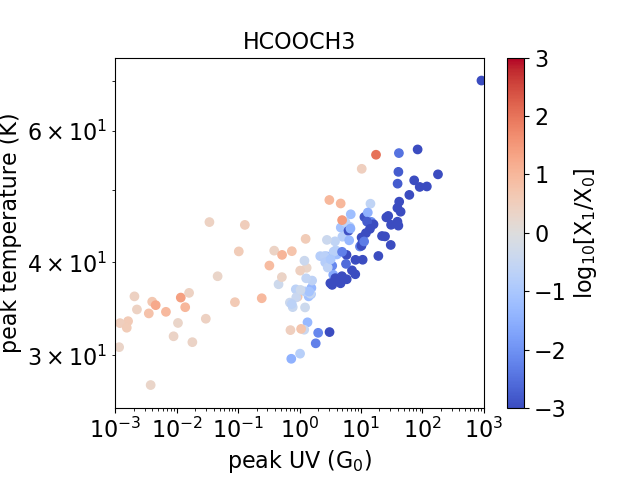}
\end{minipage}
\begin{minipage}[t]{0.3\textwidth}
 \includegraphics[width=5.5cm]{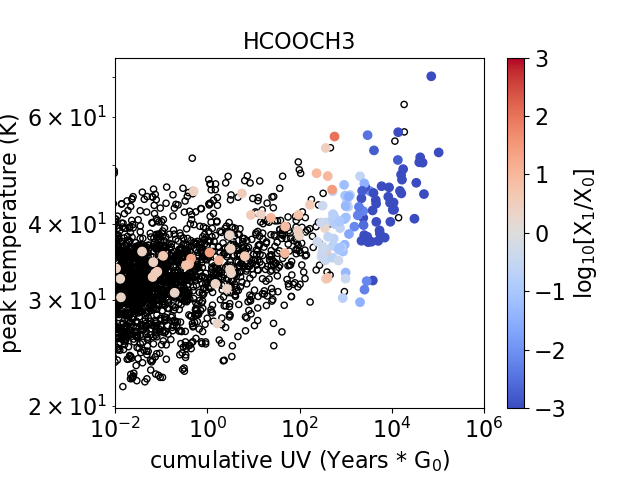}
\end{minipage} \\
\begin{minipage}[t]{0.3\textwidth}
 \includegraphics[width=5.5cm]{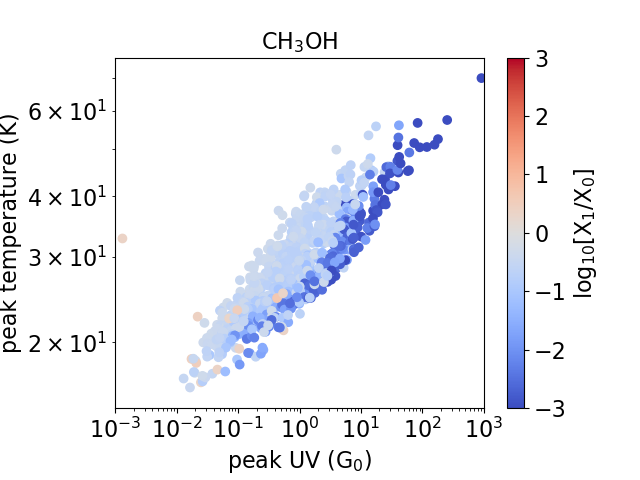}
\end{minipage}
\begin{minipage}[t]{0.3\textwidth}
 \includegraphics[width=5.5cm]{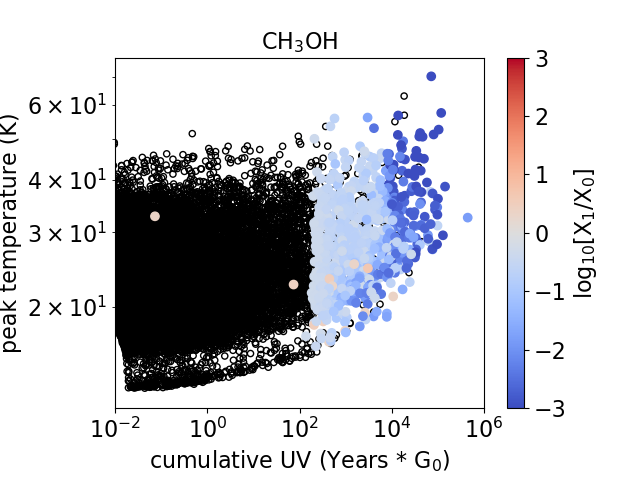}
\end{minipage} \\
\begin{minipage}[t]{0.3\textwidth}
 \includegraphics[width=5.5cm]{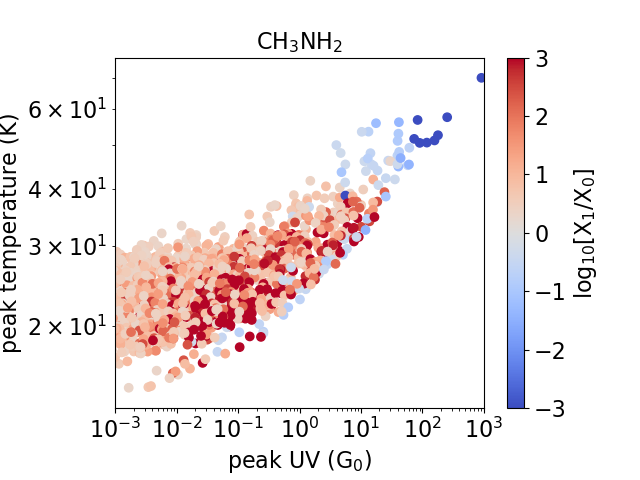}
\end{minipage}
\begin{minipage}[t]{0.3\textwidth}
 \includegraphics[width=5.5cm]{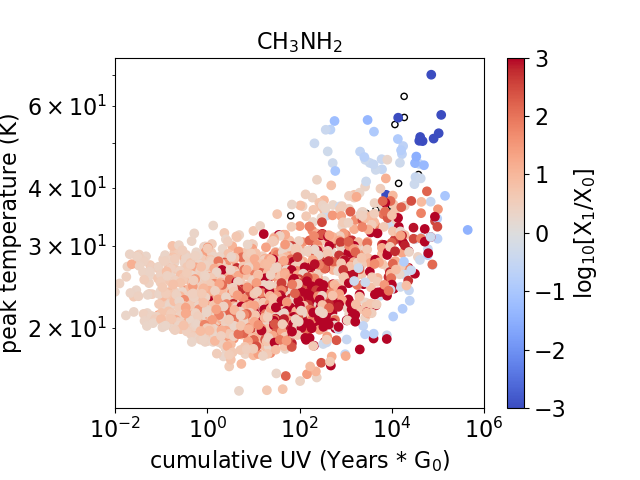}
\end{minipage}
\caption{Left panels) Peak UV flux versus peak dust temperature for the UV events relevant to CH$_3$OCH$_3$, HCOOCH$_3$, CH$_3$OH, and CH$_3$NH$_2$. The color scale indicates the logarithm of the abundance ratio \(\log_{10}(X_1/X_0)\); the red points signify an increase in abundance during the UV Event, whereas the blue points indicate a decrease. Note that the number of UV events varies among species, since we select the events relevant to their abundance change. 
Right panels) UV fluence during each UV event ($\Delta t$) versus peak dust temperature for the UV events. The open circles represent UV events where the change in molecular abundance does not meet the factor of two threshold.}
\label{fig:UV-T-X}
\end{figure*}


The left panels of Figure~\ref{fig:UV-T-X} show peak UV flux versus peak dust temperature for the UV events relevant to CH$_3$OCH$_3$, HCOOCH$_3$, CH$_3$OH, and CH$_3$NH$_2$.
The color denotes the abundance change over each UV event. It should be noted that the total number of relevant UV events varies among molecules, because of conditions (I) and (II). For example, the number of UV events plotted in the panel for HCOOCH$_3$ is much smaller than in other panels, since its abundance is often below $1\times 10^{-14}$. 
The abundances of CH$_3$OCH$_3$, HCOOCH$_3$, and CH$_3$OH exhibit a notable change with increasing peak UV flux. Initially, their abundances increase with increasing the peak UV flux but eventually begin to decrease, indicating the existence of a threshold value of the peak UV flux above which the destruction of the COMs becomes significant. Interestingly, this threshold is more clearly identified when considering the UV fluence during each UV event rather than the peak UV flux, which is shown in the right panels of Figure~\ref{fig:UV-T-X}.
The abundances of the O-bearing COMs decrease during the UV events with the UV fluence of $> 10^2$ [Years G$_0$].

The threshold value of the UV fluence for CH$_3$OH can be explained as follows. Since CH$_3$OH ice is assumed to be abundant in the initial condition, its abundance decreases in most of the UV events by photodissociation. The abundance change of CH$_3$OH ice may be described by
\begin{align}
&X_1[\text{CH}_3\text{OH}] = X_0[\text{CH}_3\text{OH}] - \int R_{\text{phdiss}}(t) dt \nonumber \\
&= X_0[\text{CH}_3\text{OH}] - X[\text{Gr}]\sigma_{\text{gr}} f^{(m)}_{\text{CH}_3\text{OH}} P_{\text{abs},\,\text{CH}_3\text{OH}} N_{\text{layer}} \int F_{\text{UV}} dt.
\end{align}
\( R_{\text{phdiss}} \) is given by equation \ref{eq:phdiss_m} with the grain number density replaced by the grain abundance. 
By substituting representative values, we obtain
\begin{equation}
X[\text{CH}_3\text{OH}] \approx X_0[\text{CH}_3\text{OH}] - 7.0 \times 10^{-8} \int F_{\text{UV}} dt,
\end{equation}
where the integral $\int F_{\rm UV}dt$ corresponds to the UV fluence in units of [$G_0$ Years] in each UV event. Here, $f^{(m)}_{\text{CH}_3\text{OH}}$, representing the fraction of CH$_3$OH in the initial ice mantle composition relative to the sum of H$_2$O, CO, and CO$_2$, is set to 0.03. In addition, $N_{\rm layer}$ is set to 400, assuming the number of monolayers formed at temperatures below 20 K. These values are assumed to be time-independent for simplicity. With an initial abundance \( X_0[\text{CH}_3\text{OH}] \) of \( 5 \times 10^{-6} \), this equation indicates that CH$_3$OH decreases significantly when the cumulative UV flux of a UV event exceeds  $\sim 100 $ Yr G$_0$.

The reduction in the CH$_3$OH abundance leads to the decreased formation rate of CH$_3$ and OH radicals, which are essential for the formation of various O-bearing COMs. When CH$_3$OH and its photodissociation products are photodissociated to form CO, it can be converted back to CH$_3$OH through the hydrogenation addition reactions. However, the hydrogen addition reactions of CO and H$_2$CO have a 2320~K activation energy barrier, making the reformation of CH$_3$OH less efficient than the photodissociation. 
In contrast, the parental molecule for N-bearing COMs is NH$_3$, where the hydrogenation addition reactions to form NH$_3$ from atomic N are barrierless, making the reformation of NH$_3$ more efficient compared to that of CH$_3$OH. As a result, a reduction in the abundance of O-bearing molecules occurs, while a reduction in the abundance of N-bearing molecules is less severe.
As we discussed earlier, nitrogen tends to exist in not only NH$_3$ but also HCN in the ice, even under high UV fluence.
As a result, CH$_3$NH$_2$ abundance increases even when the UV intensity exceeds the critical threshold.


Finally, open circles in Figure~\ref{fig:UV-T-X} represent UV events that met the criteria for the UV flux but showed less than a factor of two changes in molecular abundance during the UV event. For CH$_3$OH, it is evident that below the threshold UV fluence, the abundance change is less than a factor of two during each UV event.
Below the threshold UV fluence, CH$_3$OCH$_3$ and HCOOCH$_3$ can be formed and their abundances are enhanced.
However, there are events where no significant change in the abundances of CH$_3$OCH$_3$ and HCOOCH$_3$ even below the threshold UV fluence. 
These events happen at radii $\gtrsim$~80~au and $\lesssim$~30~au. In the outer cold region of $\gtrsim$~80~au, the peak dust temperature in the UV event is not high enough to enhance the COM formation. At the radii of $\lesssim$~30~au, on the other hand, the dust temperature is $\gtrsim$~20~K even in the midplane. Efficient formation of COMs with induced UV is possible, limiting the abundance increase during the UV events within a factor of two.

\subsection{Model uncertainties}
\subsubsection{UV attenuation in ice mantles} \label{sec:photodiss}
\begin{figure*}
 \centering
\begin{minipage}[t]{0.3\textwidth}
 \includegraphics[width=5.5cm]{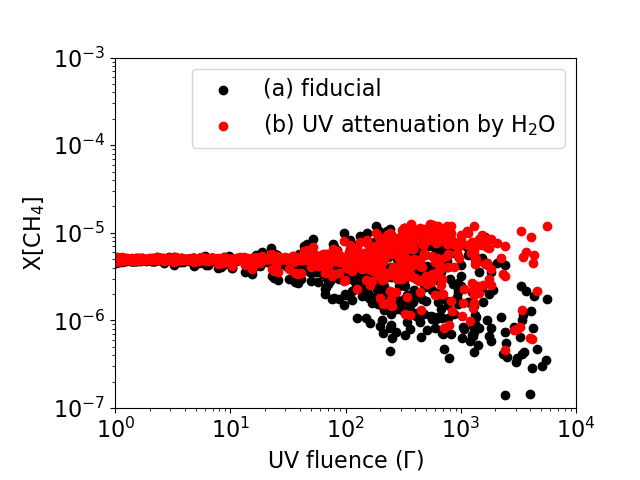}
\end{minipage}
\begin{minipage}[t]{0.3\textwidth}
 \includegraphics[width=5.5cm]{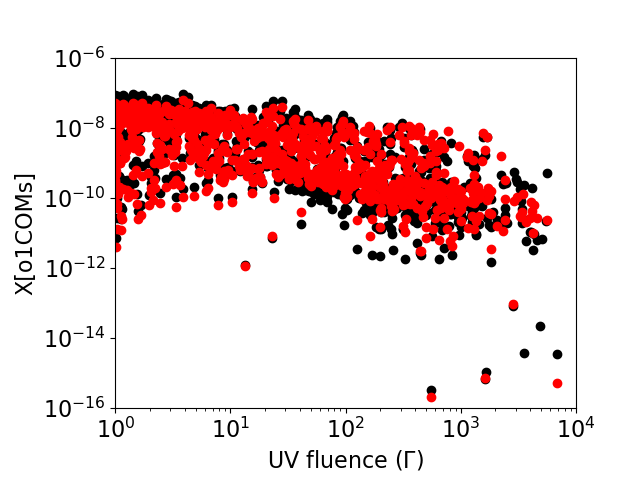}
\end{minipage}
\begin{minipage}[t]{0.3\textwidth}
 \includegraphics[width=5.5cm]{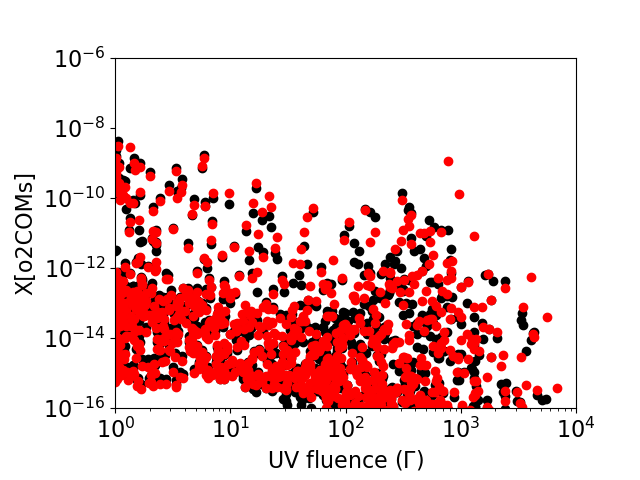}
\end{minipage}
\begin{minipage}[t]{0.3\textwidth}
 \includegraphics[width=5.5cm]{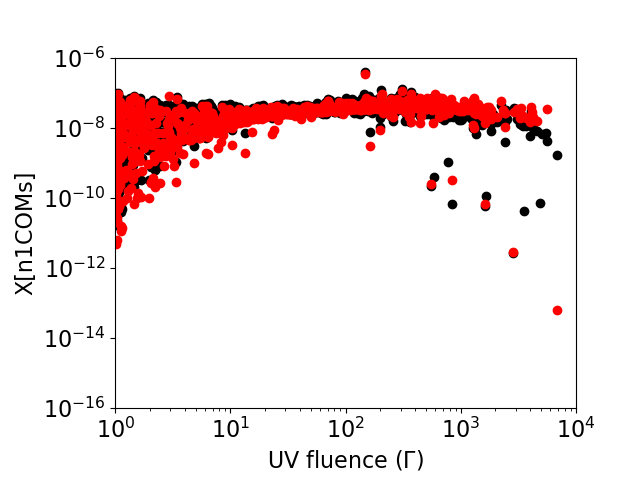}
\end{minipage}
\begin{minipage}[t]{0.3\textwidth}
 \includegraphics[width=5.5cm]{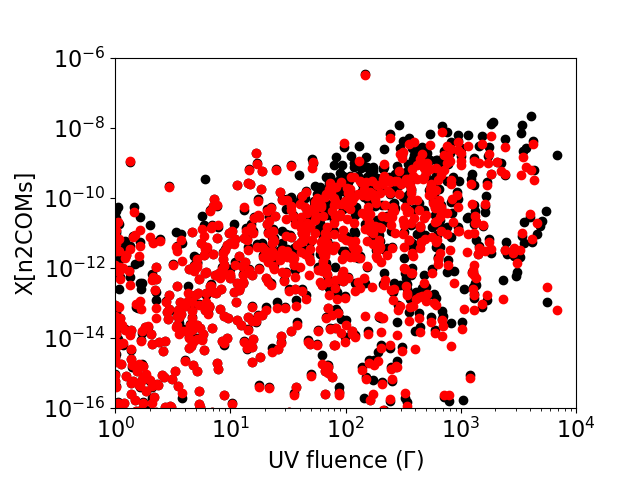}
\end{minipage}
\caption{Final abundances of CH$_4$ and COMs as functions of $\Gamma$ in the model with the UV attenuation in ice mantles by all species. For comparisons, the corresponding results in the fiducial model are shown by black dots.
}
\label{fig:ICE_compare}
\end{figure*}
%

In our fiducial model, we calculate the UV attenuation in ice mantles, assuming that species $i$ does not shield other species and vice versa.
This would correspond to assuming that the overlap of photoabsorption bands between different species is negligible.
%
%
In this case, even in the deeper layers, molecules with smaller abundances are still effectively dissociated by UV radiation.
Here we consider an alternative, extreme case; photoabsorption bands of different species completely overlap each other.
In this case, even species with low abundances can be shielded against UV radiation by other abundant species (e.g., H$_2$O) in the upper layers.
We run an additional model where the photodissociation rate of species $i$ in the ice mantle is calculated by
\begin{align}
R^{(m)}_{\rm phdiss,\,i} &= n_{\text{gr}}F_{\rm UV} \sigma_{\rm gr} f^{(m)}_i  P_{\text{abs,} i}\times 
\Sigma_{j=1}^{N_{\rm layer}}(1-\Sigma_i f^{(m)}_i P_{\text{abs,} i})^{j-1}, \label{eq:phdiss_m_overlap}
\end{align}
instead of Eq. \ref{eq:phdiss_m}.

Figure~\ref{fig:ICE_compare} compares the final abundances of selected species as functions of $\Gamma$ in the model with Eq. \ref{eq:phdiss_m_overlap} (red dots) with those in the fiducial model (black dots). CH$_4$, which is an initially abundant species, predominantly undergoes destruction rather than formation during the simulations. The abundance of CH$_4$ in the model with Eq. \ref{eq:phdiss_m_overlap} tends to be higher than that in the fiducial model; the CH$_4$ abundance averaged over all particles in the model with Eq. \ref{eq:phdiss_m_overlap} is around two times higher than that in the fiducial model.
On the other hand, the difference in the abundance of COMs between the two models remains within a factor of two. As the formation of COMs requires the formation of radicals in ice mantles via UV photodissociation, the stronger UV attenuation inside the ice mantle reduces the formation rate of COMs, while COMs themselves are also subject to destruction via UV photodissociation. 
Since the changes in the formation and destruction rates cancel each other out, the impact on the final abundances is limited.
We can conclude the treatment of UV attenuation within the ice mantle does not significantly impact the abundances of COMs in our models.

\subsubsection{Diffusion-to-desorption activation energy ratio in ice mantles} \label{sec:ebed_ratio}
\begin{figure*}
 \centering
 \begin{minipage}[t]{0.3\textwidth}
  \centering
  \textbf{$\chi$=0.4}\\
  \includegraphics[width=5.5cm]{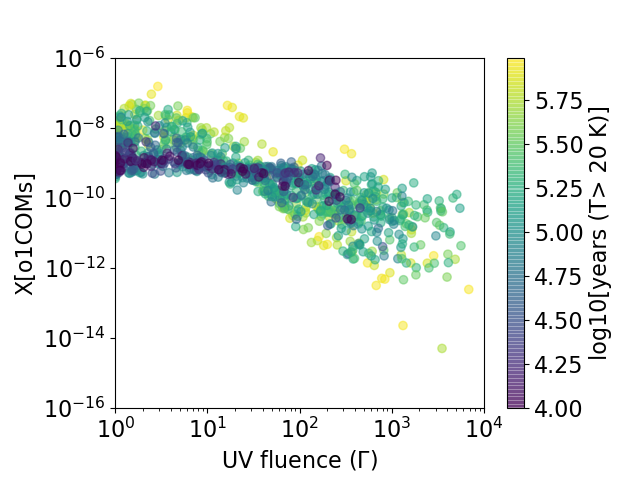}\\
  \includegraphics[width=5.5cm]{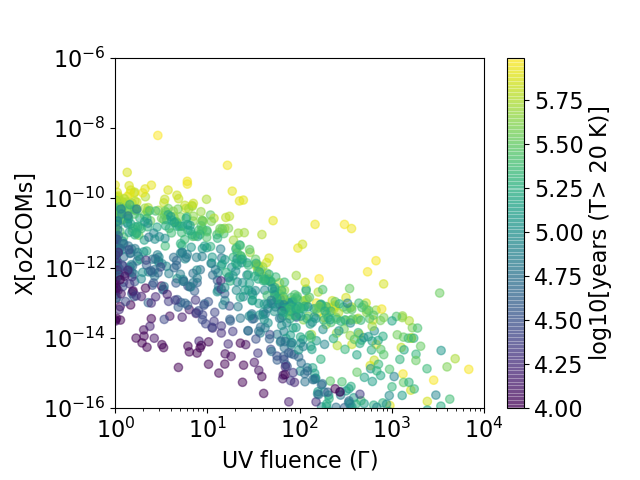}\\
  \includegraphics[width=5.5cm]{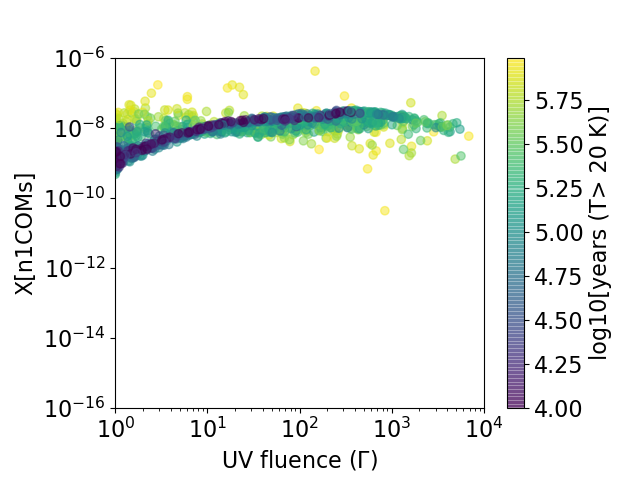}\\
  \includegraphics[width=5.5cm]{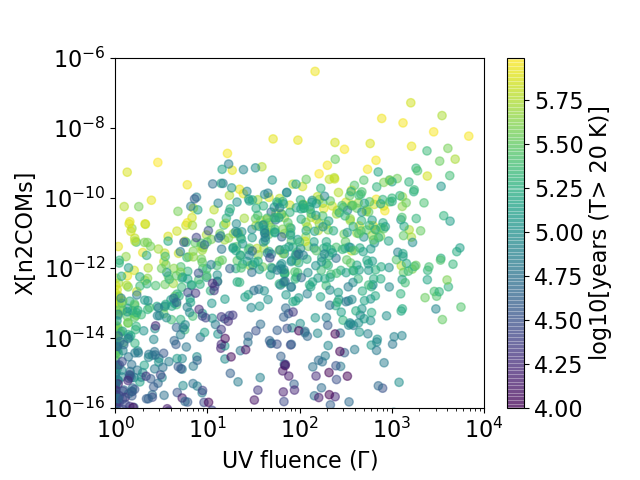}
 \end{minipage}
 \begin{minipage}[t]{0.3\textwidth}
  \centering
  \textbf{$\chi$=1.0}\\
  \includegraphics[width=5.5cm]{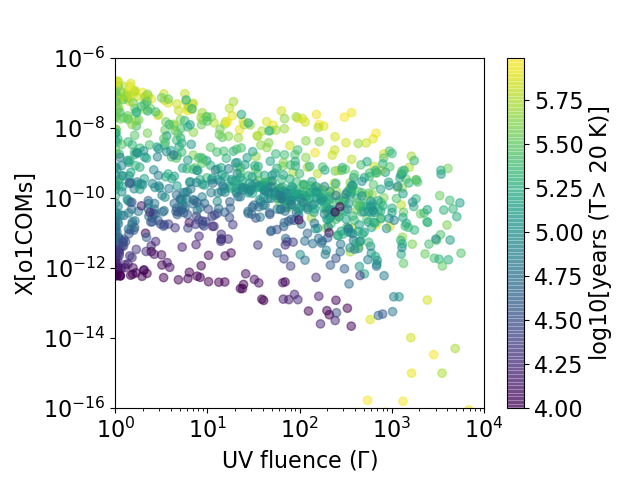}\\
  \includegraphics[width=5.5cm]{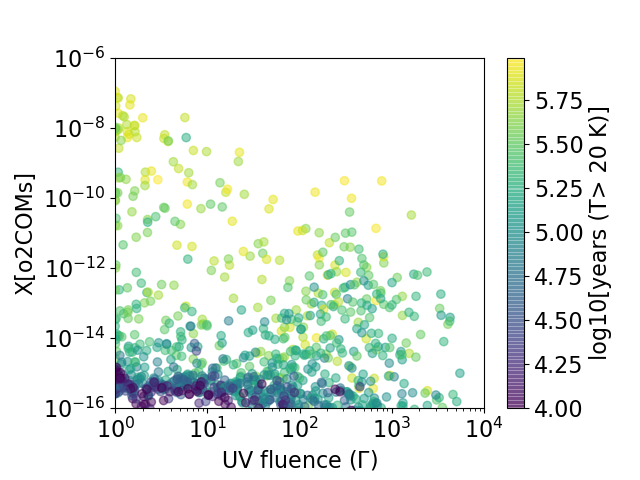}\\
  \includegraphics[width=5.5cm]{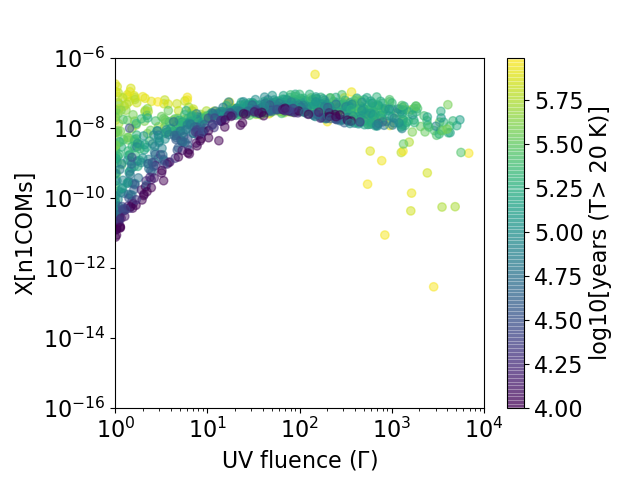}\\
  \includegraphics[width=5.5cm]{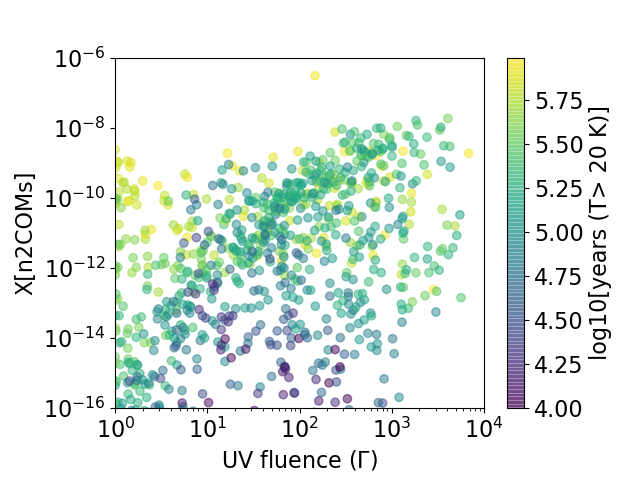}
 \end{minipage}
\caption{Final abundances of COMs as functions of $\Gamma$ in the model with $\chi$ for the bulk ice chemistry of 0.4 (left) and 1.0 (right). The colors indicate the cumulative duration when dust temperature is $\ge 20$ K.}
\label{fig:EbEdratio}
\end{figure*}

The efficiency of two-body reactions on grain surfaces and in ice mantles is determined by the activation energy for diffusion.
As commonly assumed in astrochemical models, we assume that the activation energy for diffusion is set by the desorption energy multiplied by $\chi$, where $\chi$ is the diffusion-to-desorption activation energy ratio.
In our fiducial model, $\chi$ for the ice mantle chemistry ($\chi_m$) is set to 0.8, while $\chi$ for the surface chemistry is set to 0.4.
In this subsection, we discuss the impact of assumed $\chi_m$ value on the chemical evolution of COMs.
%

Figure~\ref{fig:EbEdratio} shows the abundances of COMs as functions of $\Gamma$ in the models with $\chi_m = 0.4$ and $\chi_m =1.0$.
When $\chi_m = 0.4$, the diffusion activation energy on the surface and in the ice mantle are the same. 
One may expect that COMs can be more abundant with lower $\chi_m$ due to more efficient diffusion of radicals. The abundances of COMs in $\chi_m$ = 0.4 are, however, not always higher than those in the model with $\chi_m = 1.0$.
A noticeable difference is particularly seen in the abundance of O-bearing COMs; some particles show higher abundances of O-bearing COMs in the model with $\chi_m = 1.0$ than those in the model with $\chi_m = 0.4$.
On the other hand, for N-bearing COMs, the abundance difference between the two models is limited, and CH$_3$NH$_2$ and NH$_2$-NH$_2$ remain the most abundant N-bearing COMs in the two models as in the fiducial model.

\begin{figure*}
 \centering
 \begin{minipage}[t]{0.3\textwidth}
  \centering
  \textbf{$\chi$=0.4}\\
  \includegraphics[width=5.5cm]{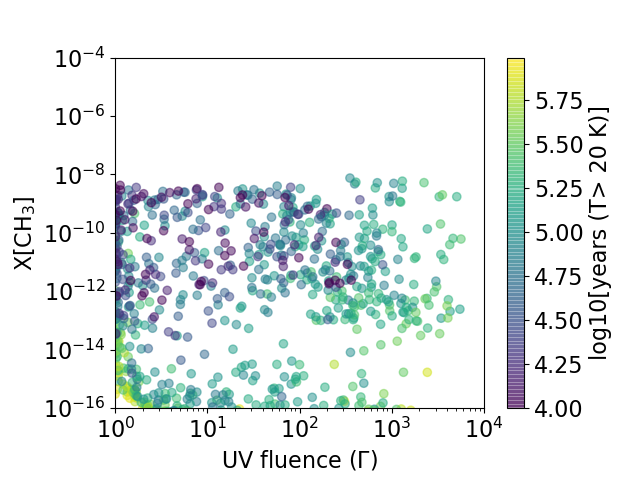}\\
  \includegraphics[width=5.5cm]{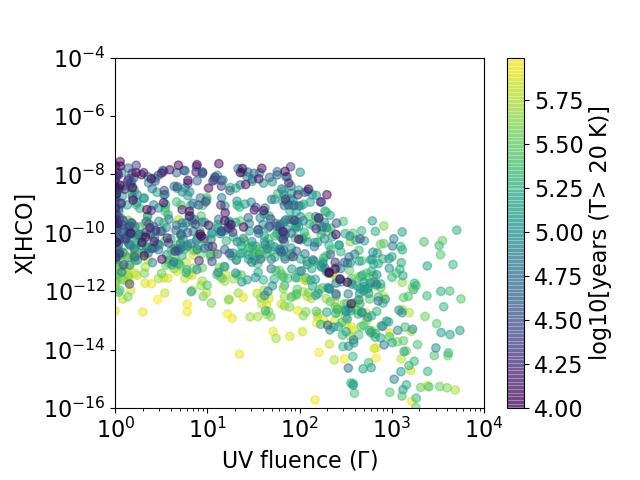}
 \end{minipage}
 \begin{minipage}[t]{0.3\textwidth}
  \centering
  \textbf{$\chi$=1.0}\\
  \includegraphics[width=5.5cm]{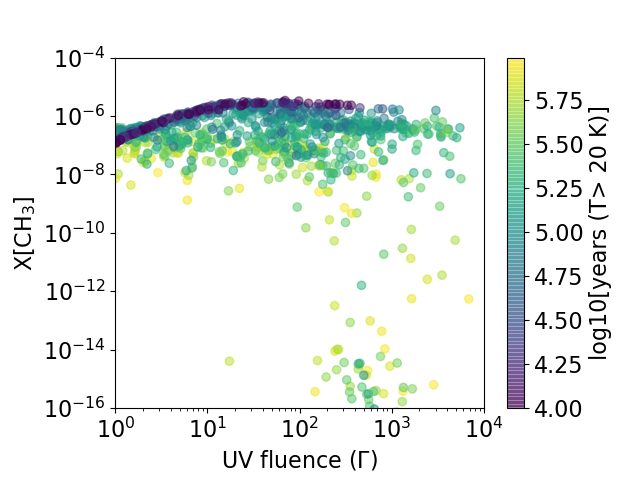}\\
  \includegraphics[width=5.5cm]{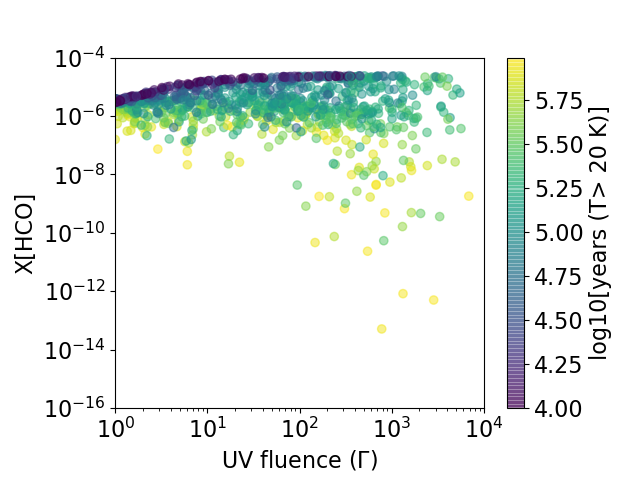}
 \end{minipage}
\caption{Final abundances of radicals on grain surface and in grain mantle, as functions of $\Gamma$ in the model with $\chi_m$ of 0.4 (left) and 1.0 (right). The colors indicate the cumulative duration when dust temperature is $\ge 20$ K.}
\label{fig:EbEdratio_radicals}
\end{figure*}

It is worth noting that the production rate of COMs is expressed as the product of the radical diffusion rate and the abundance of precursor radicals. 
Figure~\ref{fig:EbEdratio_radicals} shows the radical abundances in the models with $\chi_m = 0.4$ and with $\chi_m = 1.0$. When $\chi_m=0.4$, the diffusion of radicals in ice mantles starts at dust temperatures around 20~K, leading to the formation of COMs within the ice mantle. However, as atomic H can diffuse at even lower temperatures ($<$20 K), radicals quickly convert to stable molecules via hydrogenation, resulting in a low abundance of radicals. Consequently, the production rate of COMs when the diffusion of radicals is activated is limited by the abundance of radicals in the ice mantle.

On the other hand, when $\chi_m = 1.0$, the abundance of radicals is higher than those in the model of $\chi_m = 0.4$ by orders of magnitude, as the diffusion of atomic H is slower (Figure~\ref{fig:EbEdratio_radicals}). When the dust temperature exceeds 30~K, the diffusion of radicals in the ice mantle starts to be activated. As a result, when dust particles approach the disk surface and the temperature exceeds 30~K, the formation of COMs becomes efficient. On average, this process results in COM abundances comparable to the model with $\chi_m = 0.4$. However, the chemical evolutionary pathways differ between the models with $\chi_m = 0.4$ and $\chi_m = 1.0$. With $\chi_m = 1.0$, only radicals with low desorption energies, such as CH$_3$ and HCO, can diffuse in the ice mantles even at $\sim$30 K. Therefore, COMs related to these radicals (e.g., CH$_3$CHO, HCOOCH$_3$) becomes abundant. 
%
We also note that particles that show an increase in COMs with $\chi_m = 1.0$ are those that experience temperatures above 30~K. The temperature distribution shown in the color bar of Figure~\ref{fig:EbEdratio} indicates that particles experiencing higher temperatures are more favorable for the formation of COMs.

\subsection{Survival of COMs in the disk} \label{sec:inherit}
In the fiducial model, we assumed that initially COMs do not exist in the disk and investigated how efficiently COMs are formed in the disk. On the other hand, there is some observational evidence that supports the scenario where COMs detected in disks could be inherited from the earlier evolutionary phases \citep[e.g][]{Booth21a,Tobin23}. In this subsection, we rerun the gas-ice chemical simulations assuming that COMs are already present in the disk at $t=0$ yr (hereafter called Inherited Model) and investigate the survival of COMs in the disk.
In Inherited Model, we set the initial abundances of COMs referring to PILS survey toward the low-mass protostar IRAS 16293–2422~A \citep{Jorgensen16,Manigand20}. Among the O-bearing COMs detected by PILS survey, CH$_3$OH, HCOOCH$_3$, CH$_3$OCH$_3$, C$_2$H$_5$OH, and CH$_3$COCH$_3$ are reported to be the most abundant, while NH$_2$CHO is the most abundant among the detected N-bearing COMs. We set the initial abundance of these COMs in Inherited Model by assuming that the abundance ratio of those COMs to CH$_3$OH is the same as that obtained by PILS survey (Table~\ref{tbl:initial_abundance}).

\begin{figure*}
 \centering
 \begin{minipage}[t]{0.45\textwidth}
  \centering
  \textbf{Inherited Model}\\
  \includegraphics[width=5.5cm]{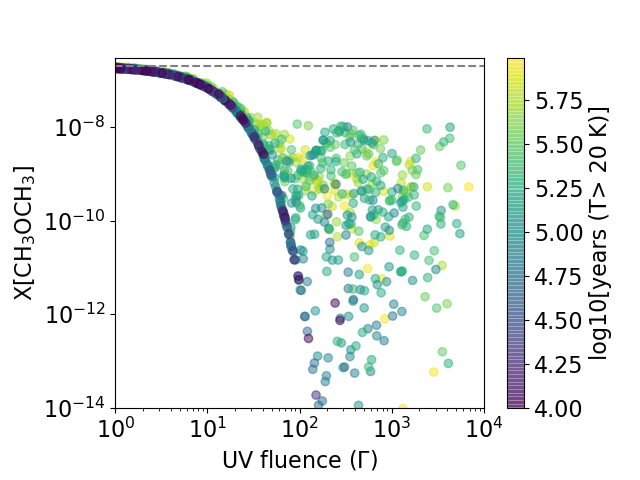}\\
  \includegraphics[width=5.5cm]{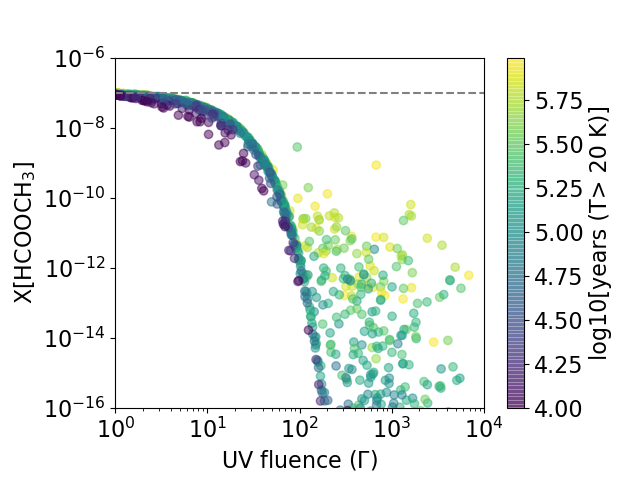}\\
  \includegraphics[width=5.5cm]{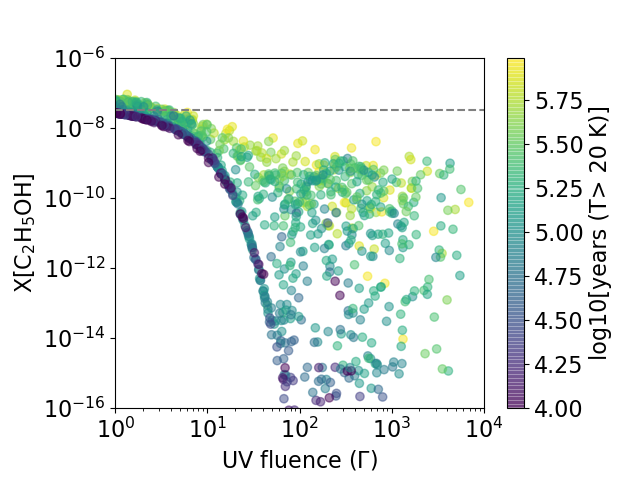}\\
  \includegraphics[width=5.5cm]{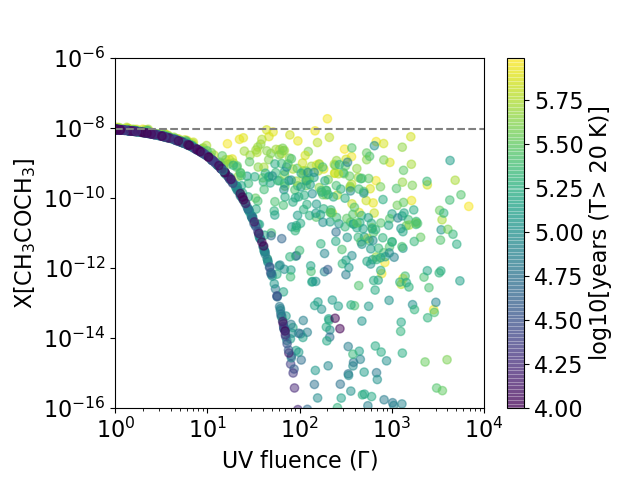}\\
  \includegraphics[width=5.5cm]{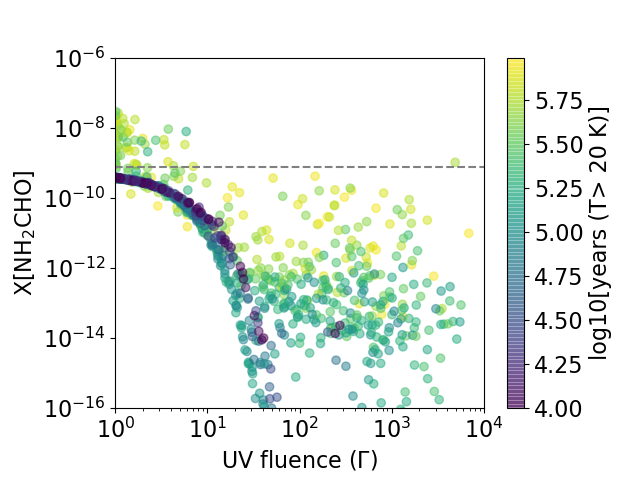}
 \end{minipage}
 \begin{minipage}[t]{0.45\textwidth}
  \centering
  \textbf{Fiducial Model}\\
  \includegraphics[width=5.5cm]{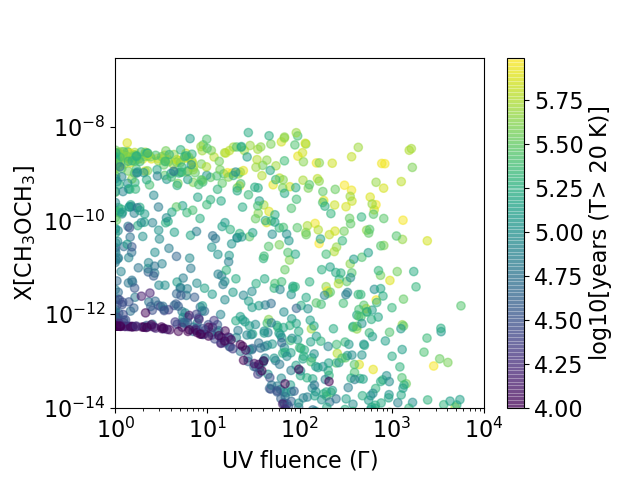}\\
  \includegraphics[width=5.5cm]{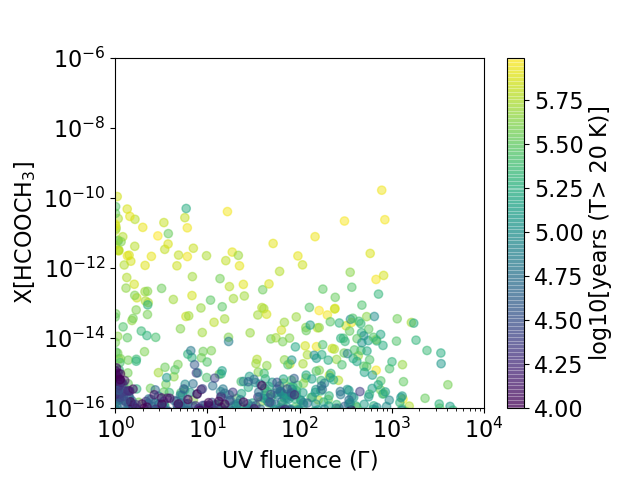}\\
  \includegraphics[width=5.5cm]{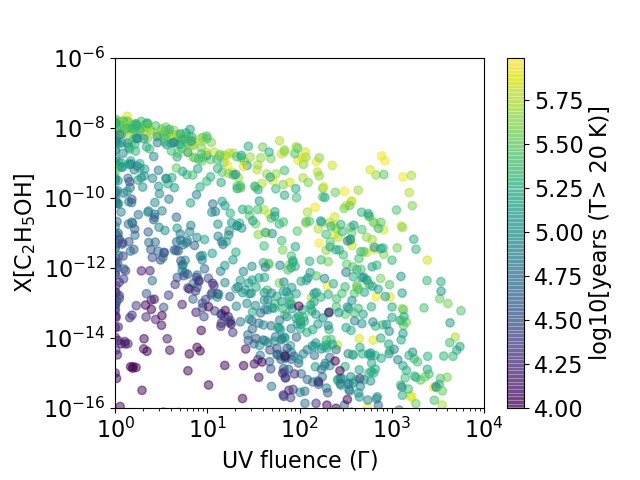}\\
  \includegraphics[width=5.5cm]{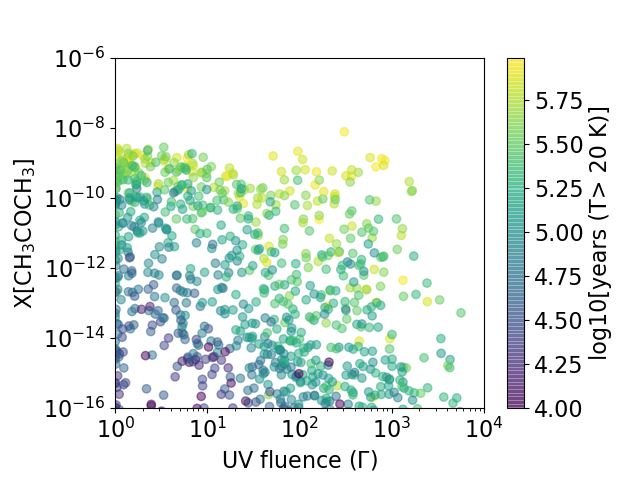}\\
  \includegraphics[width=5.5cm]{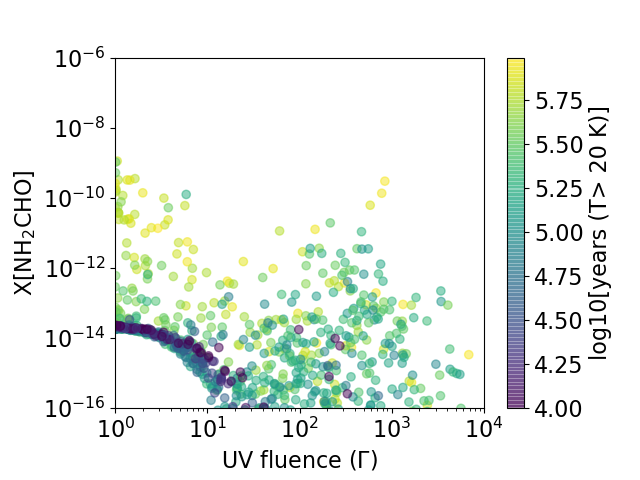}
 \end{minipage}
 \caption{Final abundances of COMs as functions of $\Gamma$ in Inherited Model (left panels) and in the fiducial model (right panels). The color scale indicates the cumulative duration when the dust temperature is above 20 K.}

 \label{fig:results_InheritedModel}
\end{figure*}

The left panels of Figure~\ref{fig:results_InheritedModel} show the abundances of COMs at $t=10^6$ yr as functions of $\Gamma$ in Inherited Model.
The initial abundances are indicated by the horizontal dashed lines in the left panels.
For comparisons, the results from the fiducial model are shown in the right panels.
For most particles, the final abundances of the five COMs are lower than their initial abundances, although the final abundances of C$_2$H$_5$OH and NH$_2$CHO are higher than their initial abundances in a small number of particles with $\Gamma \lesssim 10$.
When $\Gamma \lesssim 10$, the final COM abundances are mostly determined by the initial abundances.
When $\Gamma$ exceeds 10, both the photodissociation of COM and the formation of COMs triggered by the production of precursor radicals become efficient, leading to variations in the abundances even among particles with similar $\Gamma$ values.
Then the final COM abundances are set by both the initial COM abundances and the chemistry inside the disk.

Among COMs whose initial abundance is zero in Inherited Model, some species show higher final abundances in the Inheritance Model than in the fiducial model.
The final abundances of C$_2$H$_3$CN and C$_2$H$_5$CN averaged over all particles are ~6$\times$10$^{-10}$ and ~9$\times$10$^{-10}$, respectively.
These values are about ten times higher than in the fiducial model. 
In Inherited Model, the initial abundance of NH$_2$CHO is set to 7.7$\times$10$^{-10}$.
Through the hydrogen abstraction reaction from NH$_2$CHO, OCN is formed, followed by the grain surface reaction of OCN + C $\rightarrow$ CO + CN, providing CN radicals required for the formation of C$_2$H$_3$CN and C$_2$H$_5$CN. Therefore, the presence of NH$_2$CHO in Inherited Model leads to more efficient formation of these molecules compared to that in the fiducial model. Since C$_2$H$_3$CN and C$_2$H$_5$CN are detected in hot cores and hot corinos, our results indicate that these molecules might exist in disks as well.

%
%
%
%
%
%

\section{Conclusion}
In this work, we numerically investigated the formation and destruction of COMs in a turbulent protoplanetary disk.
%
%
We traced the motion of individual dust grains in the disk and performed gas-ice chemical network simulations along the trajectories.
%
Our models incorporated the effects of stochastic UV radiation resulting from the turbulent motion of dust grains.
We found that the stochastic UV radiation significantly influences the formation and destruction of COMs.
Our findings are summarized as follows.

\begin{enumerate}
\item Our results challenge the traditional simple assumption that the abundance of COMs is solely proportional to UV fluence. Despite similar UV fluence, the abundance of COMs can vary by orders of magnitude, due to the unique thermal and UV flux histories experienced by individual particles.
\item The abundance of O-bearing COMs is generally higher in particles with lower UV fluence. Even CR-induced UV photons alone can generate radicals, leading to the formation of O-bearing COMs. Stellar UV radiation, on the other hand, tends to photodissociate O-bearing COMs to form CO. Then CO reacts with OH radicals, which are produced by the photodissociation of H$_2$O ice, resulting in the formation of CO$_2$ ice via reaction CO + OH $\rightarrow$ CO$_2$ + H on grain surfaces. As the fraction of oxygen locked into CO$_2$ ice increases with increasing UV fluence, the abundance of O-bearing COMs decreases.
\item Our analysis of individual stellar/interstellar UV exposure events reveals that the UV fluence during each stellar UV exposure event, rather than the peak intensity of UV flux, determines whether the O-bearing COM abundances increase or decrease. 
The threshold UV fluence per UV event is $\sim$ 100 Yr G$_0$, which is sufficient for the significant reduction of CH$_3$OH.  While O-bearing COMs other than CH$_3$OH tend to increase at the UV events below the threshold fluence, the abundance change is less than a factor of two if the UV events occur inside the radius of 20~au or outside of 80~au.
In the former case, dust temperatures reach approximately 20~K even in the midplane, allowing the diffusion of radicals and thus the formation of COMs. 
Beyond 80~au from the central star, the temperature during the UV events is not high enough to proceed the formation of COMs.
\item In contrast to O-bearing COMs, high UV fluence is helpful for the formation of N-bearing COMs, as N atoms are provided by the photodissociation of N$_2$ and NH$_3$. Specifically, CH$_3$NH$_2$, a major N-bearing COM in our models, is formed by the hydrogenation of HCN, which is produced in gas-phase reactions. Unlike O-bearing COMs, the formation of the N-bearing COM can proceed even at lower temperatures, because the activation energy barrier of N atom diffusion is relatively low.
\item In calculations where the UV flux was time-averaged to maintain a consistent UV fluence, the abundances of CH$_3$OH and CH$_3$OCH$_3$ increase. This is because the UV events with the fluence higher than the threshold value is averaged out. 
On the other hand, some particles exhibited a decrease in CH$_3$NH$_2$ abundance, as its formation is related to the formation of HCN in the gas phase. It is important to note that neglecting the time-varying UV in calculations can lead to changes in molecular abundances by orders of magnitude.
\item In Inherited Model, in which the initial abundances of COMs are taken from the observations of IRAS~16293–2422~A, we found that the final abundances of the COMs are lower than the initial abundances in most particles. 
When the normalized UV fluence $\Gamma$ (Eq. \ref{eq:ft}), is less than $\sim$10, the final abundances are mostly set by the initial abundances. Conversely, at $\Gamma \gtrsim 10$, the final COM abundances are set by both the initial COM abundances and the chemistry inside the disk.
\item The UV attenuation inside ice mantles has a dual effect for the chemistry of COMs. It supplies the precursor radicals that accelerate the formation of COMs and also has the potential to destroy COMs themselves. As these effects cancel each other, UV attenuation inside ice mantles has a limited impact on the abundance of COMs, typically within a factor of two.
\item The efficiency of reactions in ice mantles is critically influenced by the activation energy for diffusion. Despite the faster diffusion expected with $\chi_m = 0.4$, where $\chi_m$ is the diffusion-to-desorption activation energy ratio in ice mantles, it is intriguing to find that the abundance of COMs is not always higher than that in the model with $\chi_m = 1.0$. Although the diffusion rate is lower in the $\chi_m = 1.0$ model, abundant radicals are stored in the ice mantle. When dust temperatures exceed 30~K, radicals such as CH$_3$ and HCO can suddenly diffuse within the mantle, driving COM formation even with $\chi_m = 1.0$. Our calculations also suggest that chemical evolutionary pathways are different in the models with $\chi_m = 0.4$ and $\chi_m = 1.0$.
\item Different impacts of UV radiation on O-bearing and N-bearing COMs are interesting results that could have insight for future observations. For example, in disks with stronger turbulence, more particles may be stirred up to the surface and to receive higher UV fluence, which could lead to a decrease in the abundance of O-bearing COMs, and potentially a decrease in the O-bearing COMs to N-bearing COMs ratio. This prediction could be tested by observing COMs in a sample of disks.
\item As a future perspective, it would be important to consider the effect of grain growth. While we thoroughly described the effect of stochastic UV irradiation on COM chemistry of migrating dust particles of radius 1~$\micron$, the grains tend to stay near the midplane when they are grown to larger size (see Figure~\ref{fig:orbits_overUV} and Appendix). Since the temporal variation of UV flux and temperature is essential for COM chemistry, it would be ideal to calculate COMs chemistry coupled with grain growth.
\end{enumerate}

\section*{Acknowledgements}
This work is supported in part by JSPS KAKENHI grant Nos. 20H05847, 21K13967, and 24K00674. Y.A. acknowledges support by NAOJ ALMA Scientific Research Grant code 2019-13B.
L.M. acknowledges the financial support of DAE and DST-SERB research grants (SRG/2021/002116 and MTR/2021/000864) from the Government of India.
Numerical computations were in part carried out on the PC cluster at the Center for Computational Astrophysics,
National Astronomical Observatory of Japan.
We thank to the anonymous referee who provided us with valuable comments.

\section*{Data Availability}

Data will be made available upon reasonable request.



\bibliographystyle{mnras}
\bibliography{COMs_in_disk} 




\appendix

\section{Results for 10 {\micron} particles}
\begin{figure*}
 \centering
\begin{minipage}[t]{0.3\textwidth}
 \includegraphics[width=5.5cm]{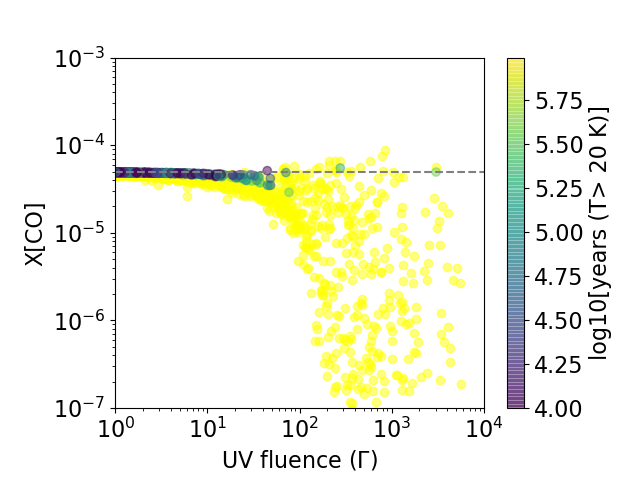}
\end{minipage}
\begin{minipage}[t]{0.3\textwidth}
 \includegraphics[width=5.5cm]{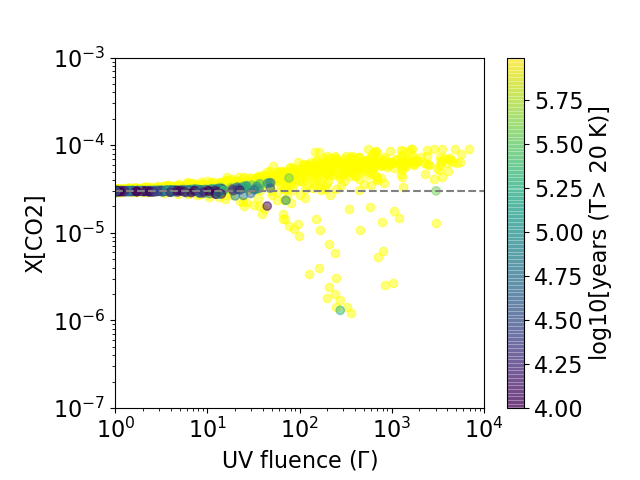}
\end{minipage}
\begin{minipage}[t]{0.3\textwidth}
 \includegraphics[width=5.5cm]{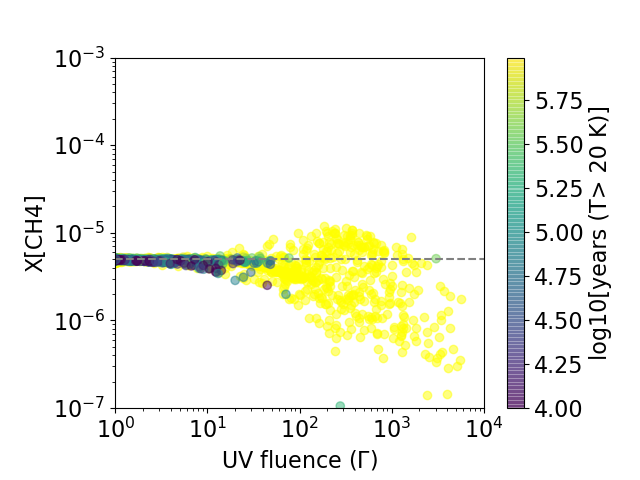}
\end{minipage}
\begin{minipage}[t]{0.3\textwidth}
 \includegraphics[width=5.5cm]{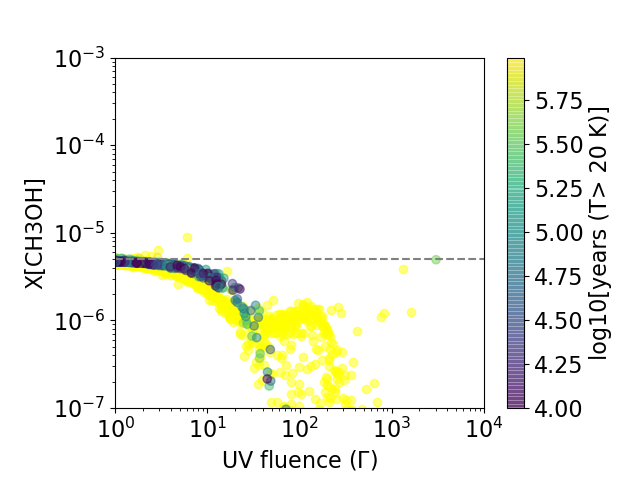}
\end{minipage}
\begin{minipage}[t]{0.3\textwidth}
 \includegraphics[width=5.5cm]{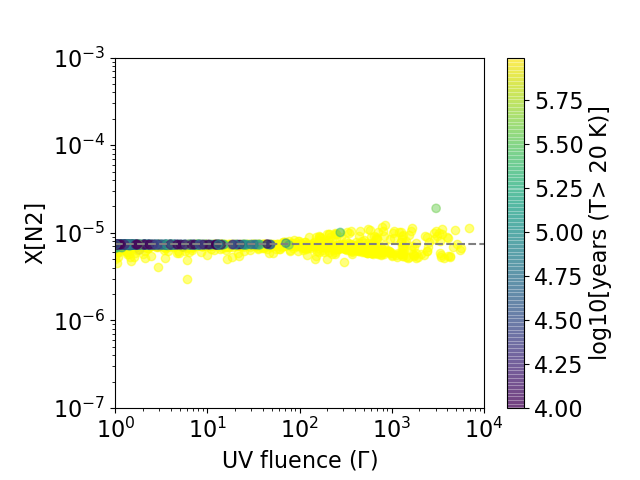}
\end{minipage}
\begin{minipage}[t]{0.3\textwidth}
 \includegraphics[width=5.5cm]{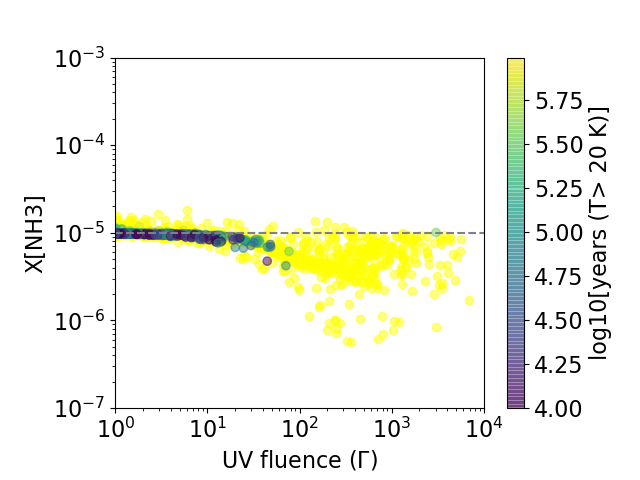}
\end{minipage}
\begin{minipage}[t]{0.3\textwidth}
 \includegraphics[width=5.5cm]{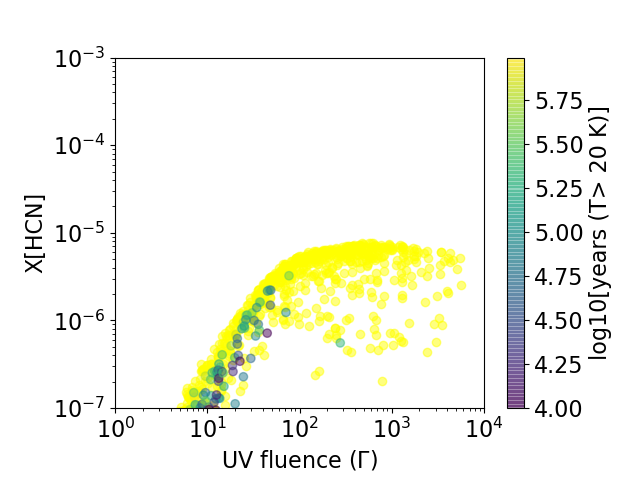}
\end{minipage}
\caption{
Similar to Fig. \ref{fig:CNO_carriers}, but for the case with the particle size of 10 \micron.
For comparisons, the yellow points represent the fiducial model where 1 \micron-sized particles are assumed.}
%
%
%
%
\label{fig:CNO_carriers_10mic}
\end{figure*}

\begin{figure*}
 \centering
\begin{minipage}[t]{0.3\textwidth}
 \includegraphics[width=5.5cm]{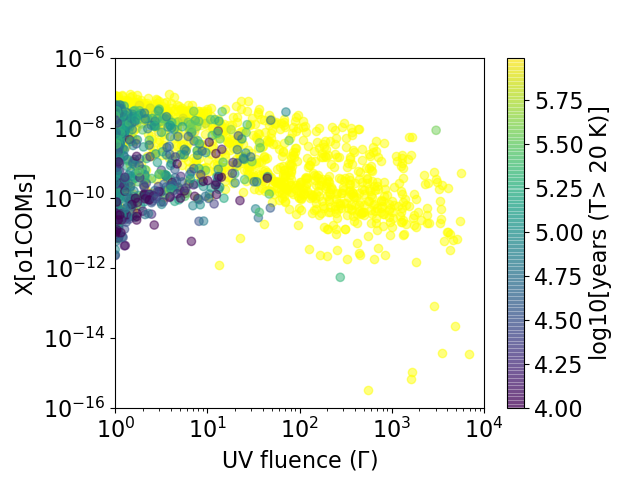}
\end{minipage}
\begin{minipage}[t]{0.3\textwidth}
 \includegraphics[width=5.5cm]{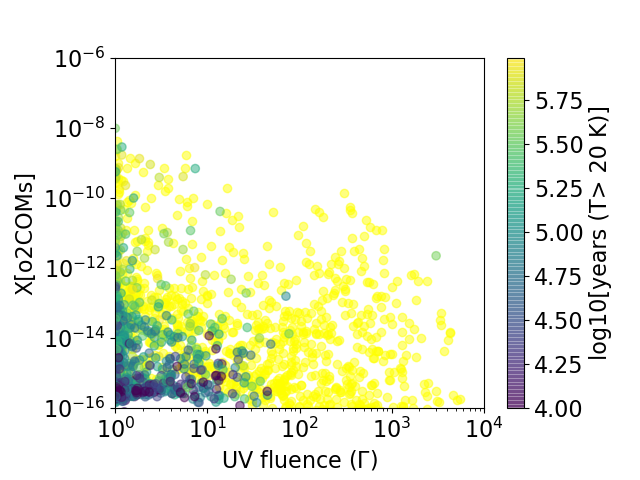}
\end{minipage} \\
\begin{minipage}[t]{0.3\textwidth}
 \includegraphics[width=5.5cm]{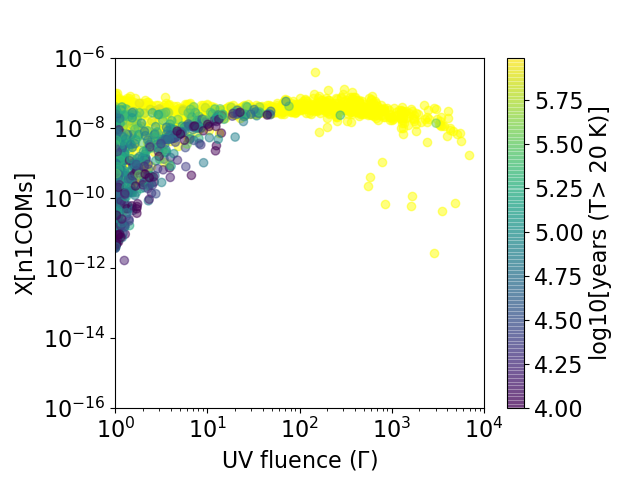}
\end{minipage}
\begin{minipage}[t]{0.3\textwidth}
 \includegraphics[width=5.5cm]{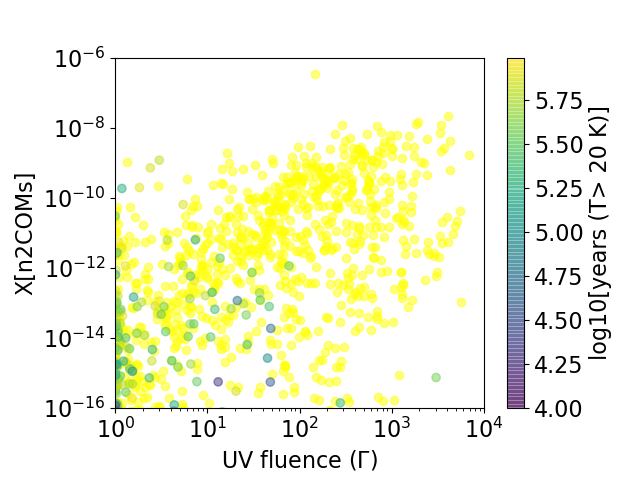}
\end{minipage}
\caption{Similar to Fig. \ref{fig:COMs}, but for the case with the particle size of 10 \micron.
For comparisons, the yellow points represent the fiducial model where 1\micron-sized particles are assumed.}
\label{fig:COMs_10mic}
\end{figure*}

In this Appendix, we discuss the effect of particle size on chemical evolution.
We rerun the particle tracking simulations and the gas-ice chemical reaction network model assuming the particle size ($a$) of 10 $\micron$ rather than 1 $\micron$.
Figure~\ref{fig:CNO_carriers_10mic} shows the results for abundant molecules, while Figure~\ref{fig:COMs_10mic} shows the results for COMs.
The yellow points in the figures represent the fiducial model, where $a =1$ $\micron$, for comparisons.
When $a =10$ $\micron$, dust particles tend to be less susceptible to turbulence, and the UV fluence is reduced by about two orders of magnitude on average compared to the fiducial model.
The conversion from CO to CO$_2$ is less efficient in the case with $a =10$ $\micron$ compared to the fiducial model even when the fluence is similar. This can be attributed to the larger proportion of mantle to the surface layer of ice when $a =10$ $\micron$, leading to slower chemical reactions.
Given the lower UV fluence, the formation of HCN is not efficient, because the formation of HCN requires the photodissociation of N$_2$ or NH$_3$.
While the production of N-bearing COMs increases with increasing the UV fluence (Section~\ref{sec:coms}), some particles still produce COMs containing one or more N, reaching abundances of 10$^{-8}$, even with $a=10$ $\micron$.
Although it was mentioned in Section \ref{sec:coms} that the production of N-containing molecules increases with increasing the UV fluence, some particles still produce COMs containing one or more N, reaching abundances of 10$^{-8}$, even with $a=10$ $\micron$.
These trends are consistent with the trend for particles with low UV fluence in Figures~\ref{fig:CNO_carriers} and \ref{fig:COMs}.

\bsp	
\label{lastpage}
\end{document}